\documentclass{SciPost}
\usepackage{tikz}

\usepackage{doi,url}
\usepackage{graphicx}
\usepackage{hyperref}
\allowdisplaybreaks

\newcommand{\prt}{\partial}
\newcommand{\sn}{\mathrm{sn}}
\newcommand{\cn}{\mathrm{cn}}

\begin{document}

\begin{center}{\Large \textbf{
Dispersive hydrodynamics of nonlinear polarization waves in
  two-component Bose-Einstein condensates}}\end{center}

\begin{center}
T. Congy\textsuperscript{1*}, A. M. Kamchatnov\textsuperscript{2},
N. Pavloff\textsuperscript{1}
\end{center}

\begin{center}
{\bf 1} LPTMS, CNRS, Univ. Paris-Sud, Universit\'e Paris-Saclay,
91405 Orsay, France
\\
{\bf 2} Institute of Spectroscopy, Russian Academy of Sciences, Troitsk,
Moscow, 108840, Russia
\\
* thibault.congy@u-psud.fr
\end{center}

\section*{Abstract}
{\bf   We study one dimensional mixtures of two-component Bose-Einstein
  condensates in the limit where the intra-species and inter-species
  interaction constants are very close. Near the mixing-demixing
  transition the polarization and the density dynamics decouple. We
  study the nonlinear polarization waves, show that they obey a
  universal (i.e., parameter free) dynamical description, identify a
  new type of algebraic soliton, explicitly write simple wave
  solutions, and study the Gurevich-Pitaevskii problem in this
  context.
}

\vspace{10pt}
\noindent\rule{\textwidth}{1pt}
\tableofcontents\thispagestyle{fancy}
\noindent\rule{\textwidth}{1pt}
\vspace{10pt}

\section{Introduction}

As demonstrated in various physical contexts, the interplay between
dispersive and nonlinear effects can lead to a number of spectacular
phenomena as, for instance, the formation of
solitons and vortices. Bose-Einstein
condensates (BECs) display both effects: (i) dispersion which is due
to the so-called quantum pressure and (ii) nonlinear properties
due to the interaction between the condensed atoms. Already in a
pioneering paper, Bogoliubov \cite{bogoliubov} showed that the
combination of these two features yields reconstruction of the ground
state of the many-particle system, with formation of new types of
elementary excitations---Bogoliubov quasiparticles. The generalization
of the Bogoliubov method to nonuniform time-dependent systems by Gross
\cite{gross} and Pitaevskii \cite{pit-1961} permitted to develop the
theory of quantum vortices and later Tsuzuki \cite{tsuzuki}
demonstrated the existence of dark solitons in a one dimensional model
of weakly interacting bosons. After the experimental realization of
BEC in ultracold gases, dark solitons were observed first in a
one-dimensional geometry under the form of dips propagating along a
stationary background \cite{burger,denschlag} and then in
two dimensions under the form of stationary oblique solitons
\cite{egk-06,kp-08,kk-11} generated by the flow of an
exciton-polariton condensate past an obstacle \cite{amo,grosso}. More
complicated nonlinear wave structures were experimentally observed
\cite{hoefer-06} and interpreted as dispersive shock waves, the
description of which can be developed in the framework of Whitham's
theory of modulations of nonlinear waves \cite{hoefer-06,kgk-04} (for a recent 
review on modulation theory of nonlinear waves see, e.g.,  Ref. \cite{El16}).

The experimental realization of condensates consisting of two (or more)
species has opened the possibility of studying even richer
dynamics triggered by the additional degree(s) of freedom consisting in the
relative motion of the components. These are
new modes that can interact with each other leading, in
particular, to different types of solitons. For two-component systems these
new modes can be visualized as pertaining to two types of waves: ``density
waves'' with in-phase motion of the two components and ``polarization
waves'' with counter-phase motion of the components. In the simplest
situations these two types of excitations decouple: the first type
does not involve relative motion of the components and the second type
of waves does not affect the total density of the condensate. In the
small amplitude limit these two types of waves and the distinction
between density and polarization excitations were studied, e.g., in
Ref.~\cite{kklp-14}.

It has been recently noticed \cite{qps-16} that the polarization
dynamics can be separated from the density dynamics even for the case
of large amplitude waves if the difference between intra- and
inter-species interaction constants is small, and this observation was
applied to the theory of polarization solitons---which were denoted as
``magnetic solitons''. In the present paper we extend the method of
Ref.~\cite{qps-16} to the general case of polarization dynamics in
two-component BECs with small difference between the nonlinear
interaction constants. In section \ref{secII} we derive the general
equations of the polarization dynamics. In section \ref{secIII} we
study their traveling wave solutions that include, as a limiting case,
the soliton solution found in \cite{qps-16} and in Sec. \ref{secIV} we
study the dispersionless limit of the nonlinear polarization
waves. This forms the basis for discussing in section \ref{secV} the
evolution of initial discontinuities in the polarization distribution.
We show that such discontinuities evolve into a wave structure
consisting in a rarefaction wave separated from a dispersive shock
wave by a plateau with constant polarization and relative flow
velocity. The main characteristics of this structure are calculated
with the use of Whitham theory and are shown to compare very well with
the results of numerical simulations. The relevance of our results for
experimental studies is discussed in section \ref{sec.discussion}. Our
conclusions are summarized in Sec.~\ref{sec.conclusion} and some
technical aspects are detailed in Appendixes \ref{app.energy} and
\ref{app.TF}.

\section{Model and main equations}\label{secII}

We consider a one-dimensional BEC system described by a two-component
spinor order parameter $\Psi(x,t)=(\psi_\uparrow,\psi_\downarrow)^t$
(where the superscript $^t$ denotes the transposition) obeying the
following coupled Gross-Pitaevskii equations
\begin{equation}\label{GP}
i\hbar \partial_t \psi_{\uparrow,\downarrow} + \frac{\hbar^2}{2 m}
\partial_{x}^2\psi_{\uparrow,\downarrow}
- \bigg[g\, |\psi_{\uparrow,\downarrow}|^2
+(g-\delta g)\, |\psi_{\downarrow,\uparrow}|^2\bigg]\psi_{\uparrow,\downarrow} =0,
\end{equation}
In Eqs.~\eqref{GP}, it has been assumed that the two intra-species non
linear coefficients $g_{\uparrow\uparrow}$ and
$g_{\downarrow\downarrow}$ have the same value, denoted as $g$. For
instance, this is exactly realized in the mixture of the two hyperfine
states $|F=1,m_F=\pm 1\rangle$ of $^{23}$Na \cite{Kno11}, and, to a
good approximation, in the mixture of hyperfine states of $^{87}$Rb
considered in Ref.~\cite{hceh-11} ($|F , m_F\rangle = |1, 1\rangle$
and $|2, 2\rangle$). The inter-species coefficient
$g_{\uparrow\downarrow}$ is written as $g-\delta g$, and we assume
that
\begin{equation}\label{cond}
0 < \delta g \ll g\; .
\end{equation}
Both conditions are realized in the above presented cases of $^{23}$Na
($\delta g/g\simeq 0.07$) and $^{87}$Rb ($\delta g/g\simeq 0.01$).
The left condition is the mean-field miscibility condition of the two
species (see, e.g., Refs. \cite{PeSm,PiSt}). The right condition will be
shown later to lead to important simplifications in the dynamics of the
system.

We represent the spinor wave function as
\begin{equation}\label{madelung}
    \left(
            \begin{array}{c}
              \psi_\uparrow \\
              \psi_\downarrow \\
            \end{array}
          \right)=
          \sqrt{\rho}\, e^{i\Phi/2}
          \left(
            \begin{array}{c}
              \cos\frac{\theta}2\,e^{-i\phi/2}e^{-i\mu_\uparrow t/\hbar} \\
              \sin\frac{\theta}2\,e^{i\phi/2}e^{-i\mu_\downarrow t/\hbar}  \\
            \end{array}
          \right) \; .
\end{equation}
In this expression, $\rho(x,t)$ is the total density and $\theta(x,t)$
governs the linear densities of the two components:
$\rho_\uparrow(x,t)=|\psi_\uparrow|^2$ and
$\rho_\downarrow(x,t)=|\psi_\downarrow|^2$ (cf. Eqs.~\eqref{eq9} in the
case of a constant total density $\rho_0$). $\Phi(x,t)$ and
$\phi(x,t)$ act as potentials for the velocity fields $v_\uparrow$ and
$v_\downarrow$ of the two components. Namely
\begin{equation}
 v_{\uparrow}(x,t)=\frac{\hbar}{2m}(\Phi_x - \phi_x)\; ,\quad
v_{\downarrow}(x,t)=\frac{\hbar}{2m}(\Phi_x+ \phi_x)\; .
\end{equation}

By means of the substitution \eqref{madelung} the
Gross-Pitaevskii system \eqref{GP} is cast into the form
\begin{equation}\label{eq1}
\begin{split}
& \hbar \rho_t+\frac{\hbar^2}{2 m}
\left[\rho(\Phi_x-\phi_x\cos\theta)\right]_x=0,\\
& \hbar \Phi_t+\frac{\hbar^2}{2 m}
\left(\frac{\rho_x^2}{2\rho^2}-
\frac{\rho_{xx}}{\rho}\right)
-\frac{\hbar^2}{2 m}\frac{\cot\theta}{2\rho}(\rho\,\theta_x)_x
+\frac{\hbar^2}{4 m}(\Phi_x^2+\phi_x^2+\theta_x^2)
+(2g-\delta g)(\rho-\rho_0)=0,\\
&\hbar  \theta_t+\frac{\hbar^2}{2 m \rho}(\phi_x\, \rho\sin\theta)_x+
\frac{\hbar^2}{2 m}\Phi_x\theta_x=0,\\
&\hbar \phi_t
-\frac{\hbar^2}{2m \rho\sin\theta}(\rho\,\theta_x)_x
+\frac{\hbar^2}{2 m}\Phi_x\phi_x
-\delta g\,\rho\cos\theta=0,
\end{split}
\end{equation}
where it is assumed that at equilibrium both components are at rest
and both have the same uniform density. The total density is denoted
as $\rho_0$. In this case the chemical potentials take
  the same value: $\mu_\uparrow=\mu_{\downarrow}=(g-\delta
  g)\rho_0/2$.  As is known, in such a system there are two types of
waves that can be called ``density'' and ``polarization'' waves. In
the small amplitude and long wavelength limit the velocity of
polarization waves that correspond to the (mainly) relative motion of
the components is equal to
\begin{equation}\label{eq2}
    c_p=\sqrt{\frac{\rho_0\delta g}{2 m}}.
\end{equation}
In the limit \eqref{cond} $c_p$ is very small compared to the
long wavelength velocity $c_d$ of density waves [$m c_d^2=\rho_0
(g-\delta g/2)$].  Following Ref.~\cite{qps-16}, we introduce also the
``polarization healing length''
\begin{equation}\label{eq3}
    \xi_p=\frac{\hbar}{\sqrt{2 m \rho_0 \delta g}}.
\end{equation}
Then the characteristic time scale for the polarization dynamics can
be measured in units of
\begin{equation}\label{eq4}
    T_p=\frac{\xi_p}{c_p}=\frac{\hbar}{\rho_0\delta g}.
\end{equation}
$T_p$ and $\xi_p$ are much larger than the corresponding
characteristic time and length associated with density waves, and for
the study of the polarization nonlinear waves it is thus appropriate to
pass to the non-dimensional variables
\begin{equation}\label{eq5}
    \zeta=\frac{x}{\xi_p},\qquad \tau=\frac{t}{T_p}.
\end{equation}
Then a very important consequence can be inferred from the second
equation (\ref{eq1}) that, in new non-dimensional variables, can be
written as
\begin{equation}\label{eq6}
    \frac{\rho-\rho_0}{\rho_0}=\frac{\delta g}{2\,g}\cdot\left\{
    \frac{\cot\theta}{\rho}(\rho\,\theta_\zeta)_\zeta-
    \Phi_\tau-\frac{\rho_\zeta^2}{2\rho^2}
    +\frac{\rho_{\zeta\zeta}}{\rho}
    -\frac12(\Phi_\zeta^2+\phi_\zeta^2+\theta_\zeta^2)+\frac{\rho}{\rho_0}-1
\right\}.
\end{equation}
We see that for $\theta\sim1$ at space and time scales of order
(\ref{eq3}) and (\ref{eq4}), correspondingly, the right-hand side
becomes small if $\delta g/g\ll1$. In this case we can assume at the
leading order that $\rho\approx\rho_0$, so that the polarization
hydrodynamics is decoupled from the density dynamics. This important
feature of the two-component BEC dynamics with a small difference of
the inter and intra-nonlinear constants was first indicated in
Ref.~\cite{qps-16} for the case of polarization solitons. The
appearance of the $\cot\theta$-function in the first term in the
braces shows that the condition $\delta g/g\ll1$ should be
complemented by another condition: $\theta$ should not be too close to
zero or $\pi$ so that the right-hand side of (\ref{eq6}) remains
small. Thus, in addition to (\ref{cond}), we assume also that
\begin{equation}\label{eq6b}
\mathrm{max}\left\{\theta,\pi-\theta\right\}\gg\frac{\delta g}g.
\end{equation}
This condition implies that the densities $\rho_\uparrow$ or
$\rho_\downarrow$ are not too close to $\rho_0$ or $0$, cf. Eqs.~\eqref{eq9}.

If the conditions \eqref{cond} and \eqref{eq6b} are fulfilled, then
the density and polarization dynamics are decoupled and we can study
the polarization dynamics separately assuming that
$\rho(x,t)=\rho_0=\mathrm{const}$ and disregarding the second equation
in the system \eqref{eq1}. This approximation greatly
simplifies the other equations. The first
one reduces to
\begin{equation}\label{eq7.0}
(\Phi_\zeta-\phi_\zeta\cos\theta)_\zeta=0\; .
\end{equation}
If we choose to work in a reference frame in which there is no flux of
the total density, Eq.~\eqref{eq7.0} simplifies to
\begin{equation}\label{eq7}
\Phi_\zeta=\phi_\zeta\cos\theta\; ,
\end{equation}
and $\Phi_\zeta$ can then be excluded from the remaining two equations. This
yields the system
\begin{equation}\label{eq8}
\begin{split}
& \theta_\tau+2\,\theta_\zeta\, \phi_\zeta\cos\theta
+\phi_{\zeta\zeta}\sin\theta=0,\\
& \phi_\tau-\cos\theta(1-\phi_\zeta^2)-\frac{\theta_{\zeta\zeta}}{\sin\theta}=0.
\end{split}
\end{equation}
This closed system of nonlinear equations shows that, 
for time scales of order
$T_p$ and length scales of order $\xi_p$, the polarization degree of
freedom decouples from the density degree of freedom, even in the
nonlinear regime.  All dimensional parameters have been scaled out
from Eqs.~\eqref{eq8} which thus correspond to a universal behavior
of polarization waves.
Note here that the relevant characteristic time \eqref{eq4} and length
\eqref{eq3} have been
identified for equal densities of both components (and will keep the same
value throughout the paper), but the validity of the
system \eqref{eq8}  does not rely on this assumption: it describes the
polarization dynamics in the limit (\ref{eq6b}), for a system
verifying \eqref{cond}. In this case, we see from Eq. \eqref{eq6}, that the 
ratio of the 
amplitude of density waves with respect to the one of polarization waves
is roughly of order $\delta g/g$.

The system \eqref{eq8} can be derived from the Hamilton principle
of extremal action \cite{qps-16} for a Lagrangian $\Lambda=\int
\mathcal{L}\, d\zeta$ with a Lagrangian density
\begin{equation}\label{eq8a}
\mathcal{L}=\phi_\tau \cos\theta-
\frac12\left[\theta_\zeta^2+(\phi_\zeta^2-1) \sin^2\theta\right] \; .
\end{equation}
From this expression we can write the (correctly dimensioned)
energy of the system under the form
\begin{equation}\label{ener.dim}
E=\frac12 \rho_0^2 \,\delta g \,\xi_p
\int d\zeta\, u(\zeta,\tau) \; .
\end{equation}
where
\begin{equation}\label{ener.density}
u = \phi_\tau\frac{\partial{\cal L}}{\partial \phi_\tau}
+ \theta_\tau\frac{\partial{\cal L}}{\partial \theta_\tau}-{\cal L}=
\frac12\left[\theta_\zeta^2+(\phi_\zeta^2-1) \sin^2\theta\right]
\end{equation}
is the energy density corresponding to the Lagrangian \eqref{eq8a}.
This expression coincides with the energy of ferromagnetic bodies in
dissipationless Landau-Lifshitz theory \cite{ll-1935}
with account of dispersion and uniaxial easy-plane anisotropy.

The system (\ref{eq8}) can be cast into other forms that may be more
convenient in some instances. In particular, the angle $\theta$ is
related to the density of each component by the formulas
\begin{equation}\label{eq9}
\rho_\uparrow=\frac12\rho_0(1+\cos\theta),\qquad
\rho_\downarrow=\frac12\rho_0(1-\cos\theta) ,
\end{equation}
hence
\begin{equation}\label{eq10}
w\equiv \cos\theta=\frac{\rho_\uparrow-\rho_\downarrow}{\rho_0}
\end{equation}
is the variable describing the variations of the relative density.
On the other hand,
\begin{equation}\label{eq11}
v\equiv \phi_\zeta=\frac{v_\downarrow - v_\uparrow}{2c_p}
\end{equation}
represents the non-dimensional relative velocity of the components. In
terms of the two variables $(w,v)$ which have clear physical meanings,
the system (\ref{eq8}) takes the form
\begin{equation}
\begin{split}\label{eq12}
& w_\tau-[(1-w^2)v]_\zeta=0\; ,\\
& v_\tau-[(1-v^2)w]_\zeta+
\left[\frac1{\sqrt{1-w^2}}
\left(\frac{w_\zeta}{\sqrt{1-w^2}}\right)_\zeta\right]_\zeta=0 \; .
\end{split}
\end{equation}
This system coincides with the one-dimensional version of the system derived in
the recent preprint \cite{ish-2016} for hydrodynamic description of
magnetization dynamics in ferromagnetic thin films.

For subsonic flows with velocities $|v|<1$ we can introduce a
variable $\sigma$ such that
\begin{equation}\label{eq13}
v=\cos\sigma,
\end{equation}
and then in terms of $(\theta,\,\sigma)$-variables the system of
equations of the polarization dynamics reads
\begin{equation}\label{eq14}
\begin{split}
 \theta_\tau+2\cos\theta\cdot\cos\sigma\cdot\theta_\zeta
&-\sin\theta\cdot\sin\sigma\cdot\sigma_\zeta=0,\\
 \sigma_\tau+2\cos\theta\cdot\cos\sigma\cdot\sigma_\zeta
&-\sin\theta\cdot\sin\sigma\cdot\theta_\zeta+\frac1{\sin\sigma}
\left(\frac{\theta_{\zeta\zeta}}{\sin\theta}\right)_\zeta=0.
\end{split}
\end{equation}

The importance of distinguishing subsonic from supersonic flows---an
essential assumption for being able to write the relation
\eqref{eq13}---can be
seen from the following observation: consider a
stationary uniform background characterized by a
relative density $w_0$ and a relative velocity $v_0$. Linear perturbations
of the form
\begin{equation}\nonumber
w=w_0+w',\quad v=v_0+v',\quad\mbox{where}\quad 
w'(\zeta,\tau),v'(\zeta,\tau)\propto\exp[i(k\zeta-\omega\tau)],
\end{equation}
obey the following dispersion relation:
\begin{equation}\label{eq15}
\omega=\left(2w_0v_0\pm\sqrt{(1-w_0^2)(1-v_0^2)+k^2}\right)k\; .
\end{equation}
By definition we always have $|w_0|\leq 1$, however $v_0$ can have any
value, and for $|v_0|>1$ the frequency $\omega$ is complex for small
enough wavevectors $k$. This implies a long wavelength instability of
supersonic relative motions of two-component superfluids, more
precisely for a background relative velocity $v_\downarrow-v_\uparrow$
larger than $2 c_p$. This mechanism of instability has been first
theoretically studied in Ref.~\cite{Law2001}, and the regime
\eqref{cond} we consider here corresponds to what is denoted as
``strong coupling'' in this reference.

We note here for future use that, for subsonic flows with
$w_0=\cos\theta_0$ and $v_0=\cos\sigma_0$, the dispersion relation
\eqref{eq15} can be written as
\begin{equation}\label{eq15b}
  \omega=\left(2\cos\sigma_0\cos\theta_0
\pm\sqrt{\sin^2\sigma_0\sin^2\theta_0+k^2}\right)k\;  .
\end{equation}
The long wave length behavior of the dispersion relations \eqref{eq15}
and \eqref{eq15b} is linear and corresponds to a velocity of sound in
the laboratory frame
\begin{equation}\label{vsound}
c =2w_0v_0\pm\sqrt{(1-w_0^2)(1-v_0^2)} =
2\cos\sigma_0\cos\theta_0 \pm\sin\sigma_0\sin\theta_0\; .
\end{equation}
For a uniform system in which both components have equal densities
($w_0=0$) and no relative velocity ($v_0=0$) one gets $c=\pm 1$,
i.e., going back to dimensional quantities, the speed of the
polarization sound is $c_p$ as expected.

\section{Traveling waves and solitons of polarization}\label{secIII}

In this section we consider traveling wave for which the
physical variables $\theta$ and $v$ depend on $\xi=\zeta-V\tau$ only,
$V$ being the phase velocity of the wave. In the framework of the
system (\ref{eq8}) this corresponds in making the {\it ansatz} that the
velocity potential $\phi(\zeta,\tau)$ and $\theta(\zeta,\tau)$ can be
represented as
\begin{equation}\label{pt.traveling}
\phi(\zeta,\tau)=q\zeta+\tilde{\phi}(\xi),\quad \theta=\theta(\xi).
\end{equation}
Substitution into the first equation of the system
(\ref{eq8}), multiplication by
$\sin\theta$ and integration give at once
\begin{equation}\label{tildephi}
\tilde{\phi}_\xi=-q+V\cdot\frac{B-\cos\theta}{\sin^2\theta},
\end{equation}
where $B$ is an integration constant. Substituting this expression
into the second equation of the system (\ref{eq8}) gives after simple
transformations the equation
\begin{equation}
\theta_{\xi\xi}=V^2\cdot\frac{(B-\cos\theta)(B\cos\theta-1)}{\sin^3\theta}
-\sin\theta\cos\theta+Vq\sin\theta.
\end{equation}
Multiplication by $\theta_\xi$ and integration yield the final equation for the
variable $w=\cos\theta$:
\begin{equation}\label{eq16}
w_\xi^2=-Q(w)\; , \quad\mbox{with}\quad Q(w)=w^4-2Vqw^3+(C-1)w^2
+2V(q-VB)w+V^2(1+B^2)-C,
\end{equation}
where $C$ is an integration constant. The four parameters $V,q,B,C$ can
be expressed
in terms of the four zeroes $w_1\leq w_2\leq w_3\leq w_4$ of the polynomial
\begin{equation}\label{eq17}
Q(w)=\prod_{i=1}^4(w-w_i)=w^4-s_1w^3+s_2w^2-s_3w+s_4,
\end{equation}
where the $s_i$'s are standard symmetric functions of the zeroes
$w_i$ \footnote{$s_1=\sum_i w_i$, $s_2=\sum_{i\ne j}w_iw_j$, $s_3=
  \sum_{i\ne j\ne k\ne i} w_iw_jw_k$ and $s_4=\Pi_i w_i$.
}.  In
particular, we obtain
\begin{equation}\label{eq17a}
V=\pm\frac12\left[Q(1)+Q(-1)+2\sqrt{Q(1)Q(-1)}\right]^{1/2},
\end{equation}
and
\begin{equation}\label{eq17b}
{q}=\frac{s_1}{2V}.
\end{equation}
The solution of Eq.~(\ref{eq16}) can be expressed in terms of Jacobi
elliptic functions and, without going into well-known details (see,
e.g., \cite{kamch-12}), we shall present here the final results.

The variable $w$ can oscillate between two zeroes of the polynomial
$Q(w)$ where $Q(w)\leq0$ provided these two zeroes are located in the
interval $[-1,1]$.  There are two possibilities, labeled as (A) and
(B) below.

(A) In the first case the periodic solution corresponds to oscillations
of $w$ in the interval
\begin{equation}\label{eq18}
w_1\leq w\leq w_2.
\end{equation}
The solution of Eq.~(\ref{eq16}) can be written as
\begin{equation}\label{eq19}
\xi=\int_{w_1}^w\frac{dw}{\sqrt{(w-w_1)(w_2-w)(w_3-w)(w_4-w)}} \; .
\end{equation}
To simplify the notations, we put in \eqref{eq19} (and in all
subsequent similar equations) the integration constant $\xi_0$ equal
to zero.  A standard calculation yields
\begin{equation}\label{eq20}
w=w_2-\frac{(w_2-w_1)\cn^2(W,m)}{1+\frac{w_2-w_1}{w_4-w_2}\sn^2(W,m)},
\end{equation}
where
\begin{equation}\label{eq21}
W=\sqrt{(w_3-w_1)(w_4-w_2)}\,\xi/2, \quad
m=\frac{(w_4-w_3)(w_2-w_1)}{(w_4-w_2)(w_3-w_1)},
\end{equation}
$\cn$ and $\sn$ being Jacobi elliptic functions \cite{Abram}.
The wavelength is given by
\begin{equation}\label{eq23}
L=\frac{4K(m)}{\sqrt{(w_3-w_1)(w_4-w_2)}},
\end{equation}
where $K(m)$ is the complete elliptic integral of the first kind
\cite{Abram}.  In the limit $w_3\to w_2$ ($m\to1$) the wavelength
tends to infinity and the solution (\ref{eq20}) transforms to a
soliton
\begin{equation}\label{eq24}
w=w_2-\frac{w_2-w_1}{\cosh^2W+\frac{w_2-w_1}{w_4-w_2}\sinh^2W}.
\end{equation}
This is a ``dark soliton'' for the variable $w=\cos\theta$.

The limit $m\to0$ can be reached in two ways.

(i) If $w_2\to w_1$, then we get
\begin{equation}\label{eq25}
w\cong w_2-\frac12(w_2-w_1)\cos[k(\zeta-V\tau)],\quad
\mbox{where}\quad k=\sqrt{(w_3-w_1)(w_4-w_1)}.
\end{equation}
This is a small-amplitude limit describing propagation of a harmonic wave.

(ii) If $w_4=w_3$ but $w_1\neq w_2$, then we get a nonlinear wave
represented in terms of trigonometric functions:
\begin{equation}\label{eq26}
 w=w_2-\frac{(w_2-w_1)\cos^2W}{1+\frac{w_2-w_1}{w_3-w_2}\sin^2W},
\quad\mbox{where}\quad
 W=\sqrt{(w_3-w_1)(w_3-w_2)}\,\xi/2.
\end{equation}
If we take the limit $w_2-w_1\ll w_3-w_1$ in this solution, then we
return to the small-amplitude limit (\ref{eq25}) with $w_4=w_3$. On
the other hand, if we take here the limit $w_2\to w_3=w_4$, then the
trigonometric functions in \eqref{eq26} have a small argument and can
be approximated by the first terms of their series expansions. This
yields a solution which we denote as an ``algebraic soliton'':
\begin{equation}\label{eq27}
w=w_2-\frac{w_2-w_1}{1+(w_2-w_1)^2(\zeta-V\tau)^2/4}.
\end{equation}

(B) In the second case the variable $w$ oscillates in the interval
\begin{equation}\label{eq28}
w_3\leq w\leq w_4
\end{equation}
so that instead of (\ref{eq19}) we get
\begin{equation}\label{eq29}
\xi=\int_{w_1}^w\frac{dw}{\sqrt{(w-w_1)(w-w_2)(w-w_3)(w_4-w)}}.
\end{equation}
Again, a standard calculation yields
\begin{equation}\label{eq30}
w=w_3+\frac{(w_4-w_3)\cn^2(W,m)}{1+\frac{w_4-w_3}{w_3-w_1}\sn^2(W,m)}.
\end{equation}
with the same definitions for $W$, $m$, and $L$
as in Eqs. (\ref{eq21}) and (\ref{eq23}).
In the soliton limit $w_3\to w_2$ ($m\to1$) we get
\begin{equation}\label{eq31}
w=w_2+\frac{w_4-w_2}{\cosh^2W+\frac{w_4-w_2}{w_2-w_1}\sinh^2W}.
\end{equation}
This is a ``bright soliton'' for the variable $w=\cos\theta$.

Again, the limit $m\to0$ can be reached in two ways.

(i) If $w_4\to w_3$, then we obtain a small-amplitude harmonic wave
\begin{equation}\label{eq32}
w\cong w_3+\frac12(w_4-w_3)\cos[k(\zeta-V\tau)],\quad\mbox{where}\quad
k=\sqrt{(w_3-w_1)(w_3-w_2)}.
\end{equation}
This is a small-amplitude limit describing a harmonic wave.

(ii) If $w_2=w_1$, then we obtain another nonlinear trigonometric solution,
\begin{equation}\label{eq33}
w=w_3+\frac{(w_4-w_3)\cos^2W}{1+\frac{w_4-w_3}{w_3-w_1}\sin^2W},
\quad\mbox{where}\quad
 W=\sqrt{(w_3-w_1)(w_4-w_1)}\,\xi/2.
\end{equation}
If we assume in this solution $w_4-w_3\ll w_4-w_1$, then we reproduce
the small-amplitude limit (\ref{eq32}) with $w_2=w_1$. On the other hand,
in the limit $w_3\to w_2=w_1$ we obtain the algebraic soliton solution:
\begin{equation}\label{eq34}
w=w_1+\frac{w_4-w_1}{1+(w_4-w_1)^2(\zeta-V\tau)^2/4}.
\end{equation}
This ends the general presentation of the different solutions of
Eq.~\eqref{eq16}.

It is now interesting to discuss in more detail the soliton solutions
which play a special role in the description of dispersive shock waves
(Sec.~\ref{sec.WGP}). The bright soliton solution \eqref{eq31}
corresponds to an increased number of particles in the ``up''
component:
\begin{equation}
\Delta N_\uparrow = \int dx \, (\rho_\uparrow^{\rm sol} - \rho_\uparrow^{(0)})\; ,
\end{equation}
where $\rho_\uparrow^{(0)}=\rho_0(1+w_2)/2$ is the background density
of the up component and $\rho_\uparrow^{\rm sol}(\zeta,\tau)
=\rho_0(1+w)/2$, $w(\xi)$ being given by \eqref{eq31}. One gets
\begin{equation}\label{nbre.sol}
\Delta N_\uparrow =
2\,\rho_0 \xi_p \,\mbox{arctan}\sqrt{\frac{w_4-w_2}{w_2-w_1}} \; .
\end{equation}
The soliton is characterized by the three zeros $w_1$, $w_2(=w_3)$ and
$w_4$ which relate to the physical variables $w_0$ (relative
background density of the components), $V$ (velocity of the soliton)
and $v_0$ (relative background velocity of the components) through
\begin{equation}\label{eq.w41}
w_2=w_0 \; , \quad\mbox{and} \quad
w_{4/1}=v_0(V-v_0 w_0) \pm \sqrt{(1-v_0^2)[1-(V-v_0w_0)^2]}\; .
\end{equation}
The energy of the soliton is the difference between the energy
\eqref{ener.dim} of the system in the presence and in the absence of
the soliton. It reads $E_{\rm sol}=\tfrac12 \delta g\rho_0^2\xi_p {\cal E}
=\tfrac12 \hbar\rho_0 c_p {\cal E}$
where
\begin{equation}\label{grand.pot.sol}
{\cal E}= \int
{d\zeta}\bigg[u(\xi)-\tfrac12 (v_0^2-1)(1-w_0^2)\bigg]\; ,
\end{equation}
$u(\xi)$ being here the
energy density \eqref{ener.density} computed for the distribution
\eqref{eq31}. It
is shown in Appendix \ref{app.energy} that
\begin{equation}\label{eq.ener.sol}
{\cal E}=\, 2 \, \sqrt{(w_4-w_2)(w_2-w_1)}
=\, 2\sqrt{(1-v_0^2)(1-w_0^2)-(V-2v_0w_0)^2}\; .
\end{equation}
The soliton solution found in \cite{qps-16} is reproduced from
Eqs.~(\ref{eq24}) and (\ref{eq31}) if we consider the situation where the
two components have equal background densities ($w_0=0$),
and no relative velocity ($v_0=0$). In this case, one gets from
Eq.~\eqref{eq.w41}
\begin{equation}\label{eq40}
w_2=w_3=0\; ,
\quad\mbox{and}\quad w_{4/1}=\pm \sqrt{1-V^2},
\end{equation}
that is $Q(w)=w^2(w^2-1+V^2)$ which agrees with formula (\ref{eq17a}).
As a result we obtain
\begin{equation}\label{eq41}
w=\cos\theta=\pm
\frac{\sqrt{1-V^2}}{\cosh\left[\sqrt{1-V^2}\,(\zeta-V\tau)\right]}\; ,
\end{equation}
and Eqs.~(\ref{eq9}) give the corresponding densities of each
component. From \eqref{nbre.sol} and \eqref{eq.ener.sol}, one sees
that this soliton corresponds to an increase of the number particles
of the up component $\Delta N_\uparrow = \frac{\pi}{2}\rho_0\, \xi_p$
and to an energy $E_{\rm sol}=
\hbar \rho_0 c_p \sqrt{1-V^2}$, in agreement with the
findings of Ref.~\cite{qps-16}. Note however that the existence of
polarization solitons of the form \eqref{eq24} and \eqref{eq31} is not
restricted to the condition of equal background densities $\rho_{\uparrow
  0}=\rho_{\downarrow 0}$ considered in Ref.~\cite{qps-16}.

Our approach made it possible to identify new algebraic solitons
\eqref{eq27} and \eqref{eq34} with unique properties which we now
briefly discuss. The algebraic soliton \eqref{eq34} can be obtained as
the limit $w_2(=w_3) \to w_1$ of \eqref{eq31}. It corresponds to an
increased number of ``up'' particles $\Delta N_\uparrow =\pi\rho_0
\xi_p$. At variance with the case of dark/bright solitons, once the
background parameters $w_0$ and $v_0$ are fixed, the velocity $V$ of
an algebraic soliton is not free. One finds that it is fixed to be
exactly the sound velocity \eqref{vsound}.  For an algebraic soliton,
one has $w_2\to w_1$ and thus the energy \eqref{eq.ener.sol} of such a
soliton is zero, as can be checked directly from \eqref{eq34} and
\eqref{grand.pot.sol}.

Also note that the dark/bright solitons \eqref{eq24} and \eqref{eq31}
are of a quite different nature than the one identified by Busch and
Anglin in Ref.~\cite{Bus01} and observed in Ref.~\cite{Bec08}. It can
be shown that if one considers the limit of a stationnary soliton of
type \eqref{eq31} with no pedestal ($w_0\to -1$), then one does not
reach the limit of the dark-bright solitons of Ref.~\cite{Bus01}, but
instead one obtains an algebraic soliton of the form
$\rho_{\uparrow}(\zeta,\tau) =\rho_0 \, (1+\zeta^2)^{-1}$.

\section{Dispersionless approximation and simple-waves}\label{secIV}

\subsection{Dispersionless hydrodynamics and Riemann equations}

If the velocity and density distributions $v$ and $w$ are smooth
enough, that is, if they experience little change over one
polarization healing length (\ref{eq3}), then we can neglect the
dispersion effects described by the last terms in the second equations
of the systems (\ref{eq12}) and (\ref{eq14})\footnote{In this regime,
  the dispersion relation \eqref{eq15} can be approximated by a
  straight line of slope $c$ [$c$ being the speed of sound, as given
  by \eqref{vsound}], which is legitimate when $k\ll 1$, i.e., for
  wave lengths large compared to $\xi_p$.} This corresponds to the
so-called dispersionless approximation. We shall present the
corresponding equations in two forms---for the variables $(w,v)$,
\begin{equation}\label{eq42}
    w_\tau-[(1-w^2)v]_\zeta=0,\quad v_\tau-[(1-v^2)w]_\zeta=0,
\end{equation}
and for the variables $(\theta,\sigma)$,
\begin{equation}\label{eq43}
    \begin{split}
& \theta_\tau+2\cos\theta\cdot\cos\sigma\cdot\theta_\zeta
-\sin\theta\cdot\sin\sigma\cdot\sigma_\zeta=0,\\
& \sigma_\tau+2\cos\theta\cdot\cos\sigma\cdot\sigma_\zeta
-\sin\theta\cdot\sin\sigma\cdot\theta_\zeta=0.
\end{split}
\end{equation}
These are equations of hydrodynamic type which can be studied by means of
well documented methods.

First of all, we find at once from the system (\ref{eq43}) that the variables
\begin{equation}\label{eq44}
r_1=\sigma-\theta, \quad\mbox{and}\quad r_2=\sigma+\theta
\end{equation}
satisfy the equations
\begin{equation}\label{eq45}
    \frac{\prt r_{1,2}}{\prt \tau}+
V_{1,2}(r_1,r_2)\frac{\prt r_{1,2}}{\prt \zeta}=0 ,
\end{equation}
where
\begin{equation}\label{eq46}
    V_{1,2}
=\frac32\cos r_{1,2}+\frac12\cos r_{2,1}
=2\cos\sigma\cos\theta\pm \sin\sigma\sin\theta ,
\end{equation}
or in terms of the variables $(v,w)$
\begin{equation}\label{eq47}
    V_{1,2}=2wv\pm\sqrt{(1-w^2)(1-v^2)}.
\end{equation}
The characteristic velocities $V_1$ and $V_2$ are the velocities of
propagation of small disturbances along a non-uniform background
$(\theta,\sigma)$ or $(w,v)$, correspondingly. In the case of a
uniform background $w=w_0=\cos\theta_0$, $v=v_0=\cos\sigma_0$ they
coincide with the sound velocities \eqref{vsound}. The variables
$r_{1,2}$ are called Riemann invariants, and Eqs.~(\ref{eq45}) are the
hydrodynamic equations written in the Riemann invariant form (see,
e.g., Ref. \cite{LL-6}). They have the familiar form of equations of
compressible gas dynamics written in terms of the Riemann invariants,
however the relationships between the Riemann invariants and the
physical variables are more complicated here than for a gaseous
system. Once $r_1$ and $r_2$ have been found, the physical variables
$w,\,v$ are given by
\begin{equation}\label{eq48}
    w=\cos[(r_1-r_2)/2],\quad  v=\cos[(r_1+r_2)/2].
  \end{equation}
  At this point, we have reduced the polarization hydrodynamic
  equations to the symmetric Riemann form (\ref{eq45}). We shall now
  study a special class of solutions of these equations.

\subsection{Simple wave solutions}\label{sec.simple}

In the framework of the hydrodynamic approximation a special role is
played by the so-called {\it simple wave} solutions that are
characterized by the fact that one of the Riemann invariants
(\ref{eq44}) is constant along the solution, so that the system
(\ref{eq45}) reduces to a single equation of the Hopf type.  For
example, let $r_2=r_2^0=\mathrm{const}$; then we get the equation
\begin{equation}\label{eq49}
\frac{\prt r_1}{\prt \tau}+V_1(r_1,r_2^0)\frac{\prt r_1}{\prt \zeta}=0
\end{equation}
for the variable $r_1$. This equation admits the well-known solution
\begin{equation}\label{eq50}
\zeta-V_1(r_1,r_2^0)\tau=f(r_1),
\end{equation}
where $f(r_1)$ is an arbitrary function. Equation (\ref{eq50})
determines the dependence of $r_1$ on $\zeta$ and $\tau$ in an implicit
form. The function $f(r_1)$ can be thought of as the inverse
function of the initial distribution of $r_1$ at the moment $\tau=0$,
i.e., $f^{-1}(\zeta)=r_1(\zeta,\tau=0)$. The simple wave solution
with constant Riemann invariant $r_1=r_1^0=\mathrm{const}$ can be
easily written in a similar form.

The importance of the simple wave solutions is related to the fact
that, generally speaking, a hydrodynamic solution of a typical problem
consists of different functions defined on several regions in the
$(\zeta,\tau)$-plane separated by lines of discontinuity of the fields
(here $\theta$ and $\sigma$). Along the so-called {\it weak
  discontinuities} one has discontinuities of the derivatives while the
functions remain continuous. In particular, if the fluid flow has a
boundary with adjacent quiescent fluid, then this boundary is a weak
discontinuity and the neighboring flow is described by a simple wave
solution (see, e.g., \cite{LL-6}).

A special role is played by self-similar solutions, for which $r_{1,2}$
depend on the self-similar variable $z=\zeta/\tau$ only. In
particular, such solutions appear in problems where the initial
distributions do not contain parameters with dimension of a length,
e.g., in the case of initial discontinuities with abrupt jumps of the
variables $w$ and/or $v$ ($\theta$ and/or $\sigma$). The jump occurs
at some coordinate that can be taken as the origin of the
$\zeta$-coordinate frame. In this case, $r_{1,2}=r_{1,2}(z)$ and the
hydrodynamic equations (\ref{eq45}) take the form
\begin{equation}\label{eq51}
(V_1-z)\frac{dr_1}{dz}=0,\quad (V_2-z)\frac{dr_2}{dz}=0.
\end{equation}
Their solutions are evidently
\begin{equation}\label{eq52}
\begin{split}
&r_2=r_2^0=\mathrm{const},\quad\mbox{and}\quad V_1(r_1,r_2^0)=z,\\
\mbox{or}\quad
&r_1=r_1^0=\mathrm{const},\quad\mbox{and}\quad V_2(r_1^0,r_2)=z.
\end{split}
\end{equation}
These are particular cases of simple wave solutions (\ref{eq50})
with $f\equiv0$. Eqs.~(\ref{eq52}) yield for the variable $\theta$ the
distributions
\begin{equation}\label{eq53}
\begin{split}
& \theta=\pm\frac12\arccos\left(\frac23z-\frac13\cos r_2^0\right)
+\frac12r_2^0+n\pi,\\
\mbox{or}\quad
& \theta=\pm\frac12\arccos\left(\frac23z-\frac13\cos r_1^0\right)
-\frac12r_1^0+n\pi,
\end{split}
\end{equation}
where the values of the constants ($r_2^0$ or $r_1^0$ and $n\in \mathbb{Z}$)
and the signs are to be determined from the boundary conditions.

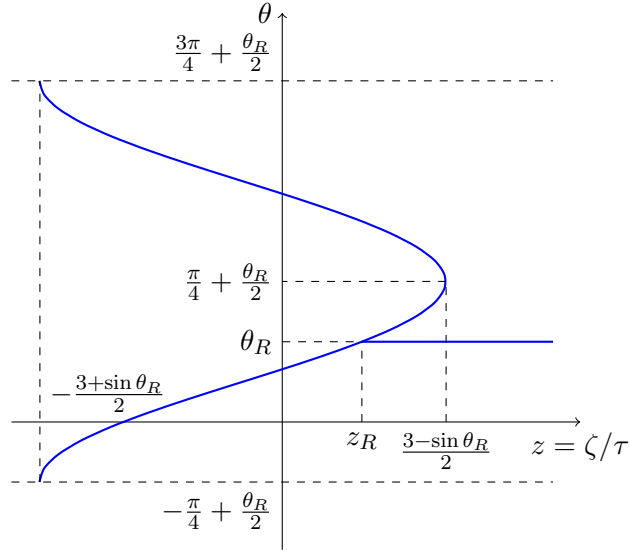
\begin{figure}
\begin{center}
\begin{tikzpicture}[xscale=1.8,yscale=1.7]
\draw [ ->] (-2.0,0.0) -- (2.2,0.0);
\node [below] at (2.2,0.0) {$z=\zeta/\tau$};
\draw[->] (0,-1) -- (0,3.2);
\node [left] at (0,3.2) {$\theta$};
\draw [dashed, - ] (-2.0,2.67) -- (2.0,2.67);
\draw [dashed, - ] (-2.0,-0.47) -- (2.0,-0.47);
\node [above left] at (0,2.67) {$\frac{3\pi}4+\frac{\theta_R}2$};
\node [below left] at (0,-0.47) {$-\frac{\pi}4+\frac{\theta_R}2$};
\draw [blue, thick, domain=-0.47:2.67] plot [samples=30, smooth] ( {1.5*sin((2*(\x)-0.628) r)-0.5*sin(0.628 r)},{(\x)});
\draw [blue, thick] (0.588,0.628) -- (2.0,0.628);
\draw [dashed, - ] (1.21,1.1) -- (0,1.1);
\node [left] at (0,1.1) {$\frac{\pi}4+\frac{\theta_R}2$};
\draw [dashed, - ] (1.21,0) -- (1.21,1.1);
\node [below] at (1.21,0) {$\frac{3-\sin\theta_R}2$};
\draw [dashed, - ] (-1.79,-0.47) -- (-1.79,2.67);
\node [above right] at (-1.79,0) {$-\frac{3+\sin\theta_R}2$};
\draw [dashed, - ] (0.588,0) -- (0.588,0.628);
\node [below] at (0.588,0) {$z_R$};
\draw [dashed, - ] (0,0.628) -- (0.588,0.628);
\node [left] at (0,0.628) {$\theta_R$};
\end{tikzpicture}
\end{center}
\caption{Distribution of $\theta(z)$ in the simple wave solution with
  fixed value of $r_2=\sigma+\theta=\pi/2+\theta_R$.  The flow is
  attached on its right side to a condensate at rest with
  $\theta=\theta_R$ and $\sigma=\pi/2$ which corresponds to the
  horizontal line. Here $z_R=\sin\theta_R$.}\label{sw-right}
\end{figure}

Let us consider here such solutions in the case where a dispersionless
polarization flow is neighboring a condensate at rest. We shall first
consider a self-similar simple wave matching at its right side a
quiescent condensate (i.e., with $\sigma=\pi/2$) where
$\theta=\theta_R=\mathrm{const}$. It is easy to see from simple
considerations \cite{LL-6} that its right edge, being a weak
discontinuity, must propagate to the right with the sound velocity
$c=\sin\theta_R$ [cf., \eqref{vsound}]; that is, this self-similar
flow has to satisfy the boundary condition $\theta=\theta_R$ at
$z=z_R=\sin\theta_R$.  Simple inspection shows that this is achieved
by the first of solutions (\ref{eq53}) (where
$r_2=\sigma+\theta=\pi/2+\theta_R=\mathrm{const}$) with a lower sign
and $n=0$. Hence, owing to the relation $\arccos x=\pi/2-\arcsin x$,
we obtain
\begin{equation}\label{eq54}
\theta=\frac12\arcsin\left(\frac23z+\frac13\sin \theta_R\right)
+\frac12\theta_R,
\end{equation}
and, consequently, by virtue of constancy of $r_2 = r_2^0 =
\frac{\pi}{2} + \theta_R$,
\begin{equation}\label{sigma-1}
  \sigma=\frac12\pi+\theta_R-\theta.
\end{equation}
It is usually supposed that $\theta$ takes values in the interval
$0\leq\theta\leq\pi$, however any interval of same length is
suitable for the description of the physical variable $w=\cos\theta$.
We shall use here the equivalent interval
\begin{equation}\label{interval-1}
  -\frac14\pi+\frac12\theta_R\leq\theta\leq\frac34\pi+\frac12\theta_R
\end{equation}
which is more suitable for the solution \eqref{eq54}. The solution
\eqref{eq54} does not cover all the interval \eqref{interval-1} over which
one has
\begin{equation}\label{zR}
  z(\theta)=\frac32\sin(2\theta-\theta_R)-\frac12\sin\theta_R\; .
\end{equation}
The resulting plot is displayed in Fig.~\ref{sw-right} for a value of
$\theta_R$ chosen in the interval $0<\theta_R<\pi$.

The left edge of this wave must have a boundary either with one of the
general solutions of equations (\ref{eq45}), or with another simple
wave with constant values of $\sigma$ and $\theta$ (that is, with a
plateau in the density distribution).  For future applications we
shall confine ourselves to the second possibility and demand that the
left edge of the solution corresponds to $\theta=\theta_0$ and,
consequently, to $\sigma=\sigma_0=\pi/2+\theta_R-\theta_0$, since
$r_2$ is constant across our simple wave. Here we have to
distinguish two main typical situations denoted as (a) and (b) below.

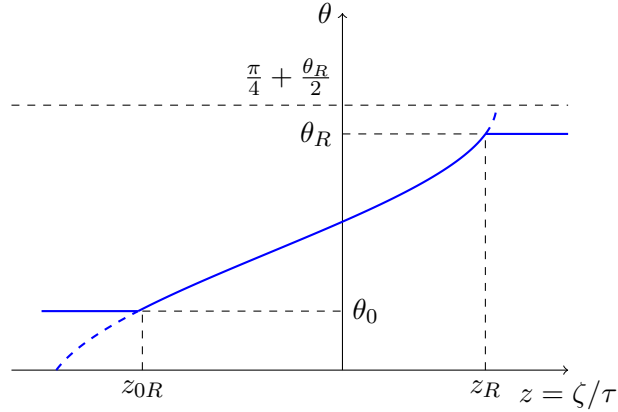
\begin{figure}
\begin{center}
\begin{tikzpicture}[xscale=2,yscale=2.5]
\draw [ ->] (-2.2,0.0) -- (1.5,0.0);
\node [below] at (1.5,0.0) {$z=\zeta/\tau$};
\draw[->] (0,0.0) -- (0,1.9);
\node [left] at (0,1.9) {$\theta$};
\draw [dashed, - ] (-2.2,1.41) -- (1.5,1.41);
\node [above left] at (0,1.41) {$\frac{\pi}4+\frac{\theta_R}2$};
\draw [blue, thick, domain=0.314:1.257] plot [samples=30, smooth] ( {1.5*sin((2*(\x)-1.257) r)-0.5*sin(1.257 r)},{(\x)});
\draw [blue, thick, dashed, domain=0:0.314] plot [samples=30, smooth] ( {1.5*sin((2*(\x)-1.257) r)-0.5*sin(1.257 r)},{(\x)});
\draw [blue, thick, dashed, domain=1.257:1.41] plot [samples=30, smooth] ( {1.5*sin((2*(\x)-1.257) r)-0.5*sin(1.257 r)},{(\x)});
\draw [blue, thick] (-2,0.314) -- (-1.36,0.314);
\draw [blue, thick] (0.95,1.257) -- (1.5,1.257);
\draw [dashed, - ] (-1.33,0.314) -- (0,0.314);
\node [right] at (0,0.314) {$\theta_0$};
\draw [dashed, - ] (0,1.257) -- (0.95,1.257);
\node [left] at (0,1.257) {$\theta_R$};
\draw [dashed, - ] (0.95,0) -- (0.95,1.257);
\node [below] at (0.95,0) {$z_R$};
\draw [dashed, - ] (-1.33,0) -- (-1.33,0.314);
\node [below] at (-1.33,0) {$z_{0R}$};
\end{tikzpicture}
\end{center}
\caption{Distribution of $\theta(z)$ in the simple wave solution; case
  (a) (see (\ref{case-1a})).  Here $z_R=\sin\theta_R$,
  $z_{0R}=\frac32\sin(2\theta_0-\theta_R)-\frac12\sin\theta_R$.}\label{fig2}
\end{figure}

Case (a): If
\begin{equation}\label{case-1a}
  -\frac14\pi+\frac12\theta_R<\theta_0<\theta_R,
\end{equation}
then the constant left flow characterized by $\sigma_0$ and $\theta_0$ is
connected with the quiescent condensate at the right by a rarefaction
wave shown in Fig.~\ref{fig2} (region $z_{0R}<z<z_R$ of this
figure) whose left edge propagates with velocity
\begin{equation}\label{velo-1}
  z_{0R}=\frac32\sin(2\theta_0-\theta_R)-\frac12\sin\theta_R.
\end{equation}

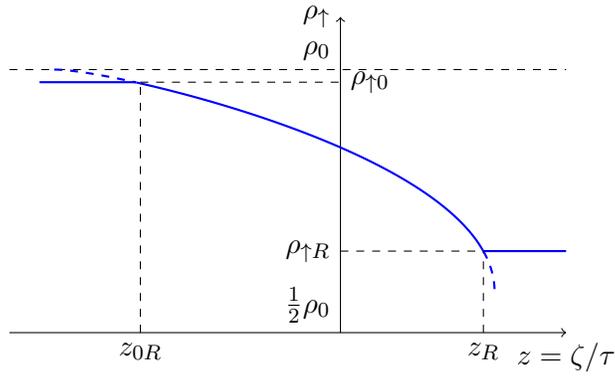
\begin{figure}
\begin{center}
\begin{tikzpicture}[xscale=2,yscale=7]
\draw [ ->] (-2.2,0.5) -- (1.5,0.5);
\node [below] at (1.5,0.5) {$z=\zeta/\tau$};
\draw[->] (0,0.5) -- (0,1.1);
\node [left] at (0,1.1) {$\rho_{\uparrow}$};
\draw [dashed, - ] (-2.2,1) -- (1.5,1.0);
\node [above left] at (0,1.0) {$\rho_0$};
\node [above left] at (0,0.5) {$\tfrac12 \rho_0$};
\draw [blue, thick, domain=0.314:1.257] plot [samples=30, smooth] ( {1.5*sin((2*(\x)-1.257) r)-0.5*sin(1.257 r)},{((1+cos(\x r))/2)});
\draw [blue, thick, dashed, domain=0:0.314] plot [samples=30, smooth] ( {1.5*sin((2*(\x)-1.257) r)-0.5*sin(1.257 r)},{((1+cos(\x r))/2)});
\draw [blue, thick, dashed, domain=1.257:1.41] plot [samples=30, smooth] ( {1.5*sin((2*(\x)-1.257) r)-0.5*sin(1.257 r)},{((1+cos(\x r))/2)});
\draw [blue, thick] (-2,0.976) -- (-1.36,0.976);
\draw [blue, thick] (0.95,0.655) -- (1.5,0.655);
\draw [dashed, - ] (-1.33,0.976) -- (0,0.976);
\node [right] at (0,0.976) {$\rho_{\uparrow 0}$};
\draw [dashed, - ] (0,0.655) -- (0.95,0.655);
\node [left] at (0,0.655) {$\rho_{\uparrow R}$};
\draw [dashed, - ] (0.95,0.5) -- (0.95,0.655);
\node [below] at (0.95,0.5) {$z_R$};
\draw [dashed, - ] (-1.33,0.5) -- (-1.33,0.976);
\node [below] at (-1.33,0.5) {$z_{0R}$};
\end{tikzpicture}
\end{center}
\caption{Distribution of $\rho_{\uparrow}(z)$ in the simple wave
  solution; case (a).  Here $z_R=\sin\theta_R$,
  $z_{0R}=\frac32\sin(2\theta_0-\theta_R)-\frac12\sin\theta_R$,
  $\rho_{\uparrow 0}=\rho_0\cos^2(\theta_0/2)$, $\rho_{\uparrow
    R}=\rho_0 \cos^2(\theta_R/2)$.}\label{fig3}
\end{figure}

\begin{figure}
\begin{center}
\begin{tikzpicture}[xscale=2,yscale=3.5]
\draw [ ->] (-2.2,0.0) -- (1.5,0.0);
\node [above] at (1.5,0.0) {$z=\zeta/\tau$};
\draw[->] (0,-1.2) -- (0,0.3);
\node [left] at (0,0.3) {$v$};
\draw [dashed, - ] (-2.2,-1) -- (1.5,-1.0);
\node [above left] at (0,0.0) {$0$};
\node [below left] at (0,-1.0) {$-1$};
\draw [blue, thick, domain=0.314:1.257] plot [samples=30, smooth] ( {1.5*sin((2*(\x)-1.257) r)-0.5*sin(1.257 r)},{(sin(((\x)-1.257) r))});
\draw [blue, thick, dashed, domain=0:0.314] plot [samples=30, smooth] ( {1.5*sin((2*(\x)-1.257) r)-0.5*sin(1.257 r)},{(sin(((\x)-1.257) r))});
\draw [blue, thick, dashed, domain=1.257:1.41] plot [samples=30, smooth] ( {1.5*sin((2*(\x)-1.257) r)-0.5*sin(1.257 r)},{(sin(((\x)-1.257) r))});
\draw [blue, thick] (-2,-0.81) -- (-1.36,-0.81);
\draw [blue, thick] (0.95,0.0) -- (1.4,0.0);
\draw [dashed, - ] (-1.33,-0.81) -- (0,-0.81);
\node [right] at (0,-0.81) {$v_0$};
\node [below right] at (0.95,0.0) {$z_R$};
\draw [dashed, - ] (-1.33,-0.81) -- (-1.33,0.0);
\node [below right] at (-1.33,0.0) {$z_{0R}$};
\end{tikzpicture}
\end{center}
\caption{Distribution of $v(z)$ in the simple wave solution; case (a).
Here
$v_0=\sin(\theta_0-\theta_R)$.}\label{fig4}
\end{figure}
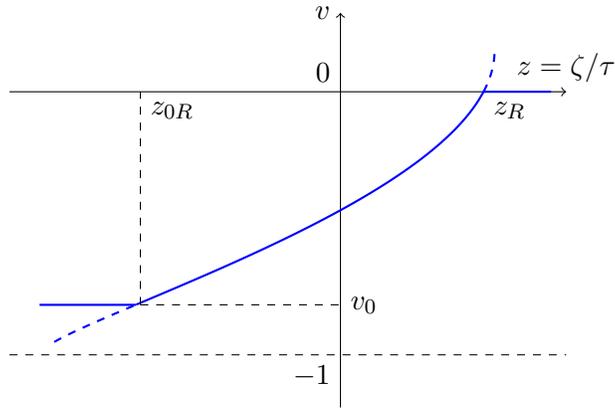

The corresponding distributions of the density $\rho_{\uparrow}$ and
the flow velocity $v=\cos\sigma=\sin(\theta-\theta_R)$ are shown in
Figs.~\ref{fig3} and \ref{fig4}, respectively.

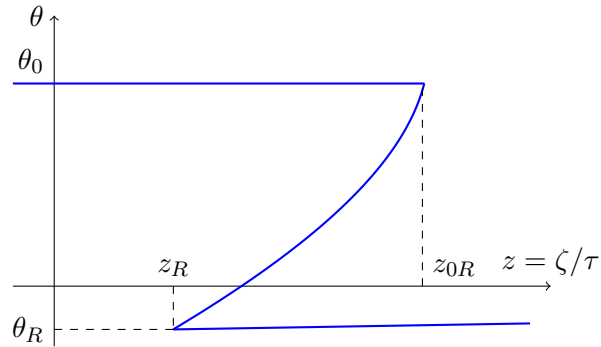
\begin{figure}
\begin{center}
\begin{tikzpicture}[xscale=5.5,yscale=8]
\draw [ ->] (0.2,0.7) -- (1.5,0.7);
\node [above] at (1.5,0.7) {$z=\zeta/\tau$};
\draw[->] (0.3,0.6) -- (0.3,1.15);
\node [left] at (0.3,1.15) {$\theta$};
\draw [blue, thick, domain=0.628:1.037] plot [samples=30, smooth] ( {1.5*sin((2*(\x)-0.628) r)-0.5*sin(0.628 r)},{(\x)});
\draw [blue, thick] (0.588,0.628) -- (1.45,0.638);
\draw [blue, thick] (0.2,1.037) -- (1.195,1.037);
\draw [dashed, - ] (0.3,0.628) -- (0.588,0.628);
\node [above left] at (0.3,1.037) {$\theta_0$};
\node [left] at (0.3,0.628) {$\theta_R$};
\draw [dashed, - ] (0.588,0.7) -- (0.588,0.628);
\draw [dashed, - ] (1.19,0.7) -- (1.19,1.037);
\node [above right] at (1.19,0.7) {$z_{0R}$};
\node [above] at (0.588,0.7) {$z_{R}$};
\end{tikzpicture}
\end{center}
\caption{Distribution of $\theta(z)$ in the simple wave solution; case
  (b1) (see (\ref{case-1c})).  Here $z_R=\sin\theta_R$,
  $z_{0R}=\frac32\sin(2\theta_0-\theta_R)-\frac12\sin\theta_R$.}\label{fig5}
\end{figure}

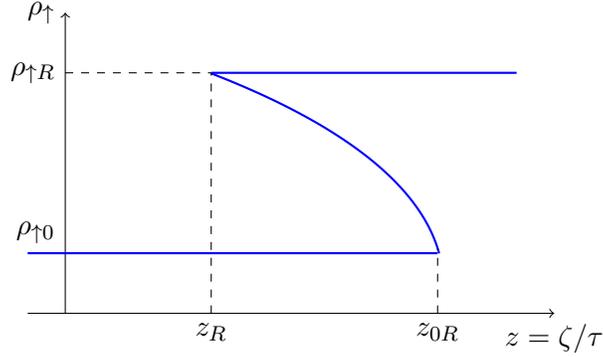
\begin{figure}
\begin{center}
\begin{tikzpicture}[xscale=5,yscale=16]
\draw [ ->] (0.1,0.7) -- (1.5,0.7);
\node [below] at (1.5,0.7) {$z=\zeta/\tau$};
\draw[->] (0.2,0.7) -- (0.2,0.95);
\node [left] at (0.2,0.95) {$\rho_{\uparrow}$};
\draw [blue, thick, domain=0.628:1.036] plot [samples=30, smooth] ( {1.5*sin((2*(\x)-0.628) r)-0.5*sin(0.628 r)},{((1+cos(\x r))/2-0.005)});
\draw [blue, thick] (0.1,0.75) -- (1.19,0.75);
\draw [blue, thick] (0.588,0.9) -- (1.4,0.9);
\draw [dashed, - ] (0.2,0.9) -- (0.588,0.9);
\node [left] at (0.2,0.9) {$\rho_{\uparrow R}$};
\node [above left] at (0.2,0.75) {$\rho_{\uparrow 0}$};
\draw [dashed, - ] (0.588,0.7) -- (0.588,0.90);
\node [below] at (0.588,0.7) {$z_R$};
\draw [dashed, - ] (1.19,0.7) -- (1.19,0.754);
\node [below] at (1.19,0.7) {$z_{0R}$};
\end{tikzpicture}
\end{center}
\caption{Distribution of $\rho_{\uparrow}(z)$ in the simple wave
  solution; case (b1).  Here $z_R=\sin\theta_R$,
  $z_{0R}=\frac32\sin(2\theta_0-\theta_R)-\frac12\sin\theta_R$.}\label{fig6}
\end{figure}

Case (b):
\begin{equation}\label{case-b1}
  \theta_R<\theta_0<\frac34\pi+\frac12\theta_R .
\end{equation}
We will see that in this case there is an interval on the $z$-axis
where the formal solution of the hydrodynamic equations becomes
three-valued. Although such a solution does not have a direct physical
meaning, it provides important relations remaining correct after
replacement of the nonphysical multi-valued parts of the flow by a
dispersive shock wave. To be definite, we illustrate such a situation
in Fig.~\ref{fig5} which is drawn in the subcase we denote as (b1) in
which
\begin{equation}\label{case-1c}
  \theta_R<\theta_0<\frac14\pi+\frac12\theta_R .
\end{equation}
In this case, in the region of the simple wave, $\theta(z)$ is given by
the single-valued solution (\ref{eq54}), but the matching with the
left and right boundaries can only be performed at the price of
overlapping the region of validity of the single wave solution with the
ones of the plateau at the boundary. This corresponds to an overall
multi-valued solution, as shown in Fig.~\ref{fig5}.
The corresponding plot of the density is shown in Fig.~\ref{fig6} and
a similar graph can be plotted for the flow velocity $v(z)$.

In the subcase we denote as (b2) for which
\begin{equation}\label{case-1d}
  \frac14\pi+\frac12\theta_R<\theta_0<\frac34\pi+\frac12\theta_R,
\end{equation}
the simple wave solution obtained from \eqref{zR} already corresponds to a
multi-valued $\theta(z)$ and the graphs of the formal hydrodynamic
solutions can be easily plotted.

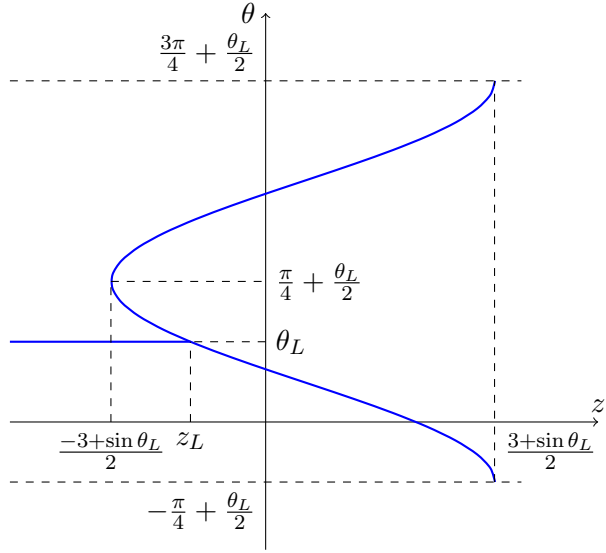
\begin{figure}
\begin{center}
\begin{tikzpicture}[xscale=1.7,yscale=1.7]
\draw [ ->] (-2.0,0.0) -- (2.6,0.0);
\node [above] at (2.6,0.0) {$z$};
\draw[->] (0,-1) -- (0,3.2);
\node [left] at (0,3.2) {$\theta$};
\draw [dashed, - ] (-2.0,2.67) -- (2.0,2.67);
\draw [dashed, - ] (-2.0,-0.47) -- (2.0,-0.47);
\node [above left] at (0,2.67) {$\frac{3\pi}4+\frac{\theta_L}2$};
\node [below left] at (0,-0.47) {$-\frac{\pi}4+\frac{\theta_L}2$};
\draw [blue, thick, domain=-0.47:2.67] plot [samples=30, smooth] ( {-1.5*sin((2*(\x)-0.628) r)+0.5*sin(0.628 r)},{(\x)});
\draw [blue, thick] (-0.588,0.628) -- (-2.0,0.628);
\draw [dashed, - ] (-1.21,1.1) -- (0,1.1);
\node [right] at (0,1.1) {$\frac{\pi}4+\frac{\theta_L}2$};
\draw [dashed, - ] (-1.21,0) -- (-1.21,1.1);
\node [below] at (-1.21,0) {$\frac{-3+\sin\theta_L}2$};
\draw [dashed, - ] (-0.588,0) -- (-0.588,0.628);
\draw [dashed, - ] (1.79,-0.47) -- (1.79,2.67);
\node [below right] at (1.79,0) {$\frac{3+\sin\theta_L}2$};
\node [below] at (-0.588,0) {$z_L$};
\draw [dashed, - ] (0,0.628) -- (-0.588,0.628);
\node [right] at (0,0.628) {$\theta_L$};
\end{tikzpicture}
\end{center}
\caption{Distribution of $\theta(z)$ in the simple wave solution with
  fixed value of $r_1=\sigma-\theta=\pi/2-\theta_L$.  The flow with
  $\theta=\theta_L$ and $\sigma=\pi/2$ (condensate at rest) can be
  attached to this solution on its left edge.  It is shown by the
  horizontal line. Here $z_L=-\sin\theta_L$.}\label{sw-left}
\end{figure}

Let us now turn to a self-similar simple wave propagating to the left
into a quiescent condensate with $\sigma=\pi/2$, $\theta = \theta_L =
\mathrm{const}$. This problem is obviously symmetric to the one just studied:
the left edge of the wave propagates here to the left with the
sound velocity $c=-\sin\theta_L$ that is, we have to satisfy the
boundary condition $\theta=\theta_L$ at $z=z_L=-\sin\theta_L$.  This
time we have to consider the second of solutions (\ref{eq53}) (where
$r_1=\sigma - \theta= \pi/2-\theta_L = \mathrm{const}$) with an upper
sign and $n=0$. Hence, we obtain
\begin{equation}\label{eq54b}
\theta=-\frac12\arcsin\left(\frac23z-\frac13\sin \theta_L\right)
+\frac12\theta_L\; ,
\end{equation}
and, consequently,
\begin{equation}\label{sigma-2}
  \sigma=\frac12\pi-\theta_L+\theta.
\end{equation}
It is clear that the plots for this case can be obtained from the
previous ones by the change $z\to-z$ replacing the notation
$\theta_R\to\theta_L$, etc.  Therefore we shall illustrate such a
situation only by the plot of $\theta(z)$ which is displayed in
Fig.~\ref{sw-left}.

Thus, we have obtained simple wave solutions which match on one
boundary with a quiescent uniform condensate, and on the other with a flow
with constant density and velocity---the ``plateau solution''.

Two important typical situations have been identified in
  this section. First, in some cases, the plateau solution can be
connected to a simple wave solution joining a quiescent condensate on
its other boundary. This is the situation illustrated in
Figs.~\ref{fig3} and \ref{fig4}.  For such flows the dispersionless
hydrodynamic approach is indeed legitimate, and it is just expected
that a more precise treatment of the weak discontinuities should
exhibit a small amount of linear radiation (on both sides of the
simple wave). Such flows are called {\it rarefaction waves}.  Second,
in some instances, the solution of the dispersionless hydrodynamic
approach is multi-valued in some regions of space,
cf. Fig.~\ref{fig6}. In these regions, the physical flow is expected
to be a {\it dispersive shock wave}, as commonly encountered in
similar situations.  In the next section we shall consider a
configuration where these two possibilities are realized.

\section{Evolution of a step-like discontinuity}\label{secV}

As a typical application of the theory, let us consider an
initial step-like distribution of polarization
\begin{equation}\label{eq55}
\theta(\zeta,\tau=0)=
\begin{cases}
 \theta_L\; , & \mbox{when} \quad \zeta<0\; ,\\
 \theta_R\; , & \mbox{when} \quad \zeta>0\; .
\end{cases}
\end{equation}
and we assume here that the left and right asymptotic regions are both
initially at rest,
\begin{equation}\label{eq56}
\sigma(\zeta,\tau=0)=
\begin{cases}
 \sigma_L=\frac{\pi}{2}\; , & \mbox{when} \quad \zeta<0\; ,\\
 \sigma_R=\frac{\pi}{2}\; , & \mbox{when} \quad \zeta>0\; .
\end{cases}
\end{equation}
We shall consider this problem in the framework of the polarization
dynamics governed by Eqs.~(\ref{eq8}), (\ref{eq12}) or
(\ref{eq14}). We shall begin with the dispersionless hydrodynamic
approximation corresponding to Eqs.~(\ref{eq42}) or (\ref{eq43}) that
can be written in the Riemann invariant form (\ref{eq45}).

\subsection{Hydrodynamic approximation}\label{hyd-appro}

The step-like discontinuity evolves into a wave whose edges propagate
into quiescent regions located at $\zeta\to\pm\infty$. If such an edge
is represented by a weak discontinuity, then the adjacent flow is
described by a simple wave solution. The step-like initial
distribution (\ref{eq55}) does not include any parameter having the
dimension of a length and, consequently, the solution has to depend
only on the self-similar variable $z=\zeta/\tau$ (and of course also,
parametrically, on $\theta_{L}$ and $\theta_R$).

One cannot find a single simple wave joining its right and left
boundaries with asymptotic regions corresponding to the initial
conditions \eqref{eq55} and \eqref{eq56}. Instead, the initial
discontinuity evolves, for $\tau>0$, into a more complex structure: an
expanding self-similar wave consisting of two simple waves separated
by a plateau characterized by the constant parameters $\theta_0$ and
$\sigma_0$. One edge of each simple wave has a boundary with a
condensates whose parameters are given by one (the left or the
right) of the boundary conditions (\ref{eq55}) and \eqref{eq56}, the
other edge matching the plateau distribution. As was discussed in the
preceding section, along the simple wave solution [matching with the
left asymptotic region $\sigma=\pi/2$, $\theta=\theta_L$] we have
$r_1=\sigma-\theta= \pi/2-\theta_L=\sigma_0-\theta_0$, and along the
other simple wave solution [matching with the right asymptotic region
$\sigma=\pi/2$, $\theta=\theta_R$] we have $r_2=\sigma+\theta=
\pi/2+\theta_R=\sigma_0+\theta_0$. These two conditions determine the
parameters of the flow on the plateau:
\begin{equation}\label{eq57}
\theta_0=\frac12(\theta_L+\theta_R),\quad
\sigma_0=\frac12(\theta_R-\theta_L+\pi).
\end{equation}
Combining with the simple wave solutions (whose characteristics are
discussed in the previous section), we find the full solution of the
problem -- determined within the dispersionless approach -- under the
form
\begin{equation}\label{eq58}
  \theta(z) =\begin{cases}
 &\theta_L,\quad z<z_L,\\
&\tfrac12\theta_L-\tfrac12\arcsin\left(\tfrac23z-\tfrac13\sin \theta_L\right),
\quad z\in(z_L,z_{0L}),\\
&\tfrac12(\theta_L+\theta_R),\quad z_{0L}<z<z_{0R},\\
&\tfrac12\theta_R+\tfrac12\arcsin\left(\tfrac23z+\tfrac13\sin \theta_R\right),
\quad z\in(z_{0R},z_R),\\
& \theta_R,\quad z>z_R,
\end{cases}
\end{equation}
where
\begin{equation}\label{eq59}
\begin{split}
& z_L= -\sin\theta_L,\\
& z_{0L}=\tfrac12\sin\theta_L -\tfrac32\sin\theta_R,\\
& z_{0R}=\tfrac32\sin\theta_L-\tfrac12\sin\theta_R,\\
&z_R=\sin\theta_R.
\end{split}
\end{equation}
The edge at $\zeta=-z_L\cdot\tau$ propagates to the left at velocity
$-\sin\theta_L$ which is the sound velocity in the left condensate.
The edge at $\zeta=z_R\cdot\tau$ propagates to the right with velocity
$\sin\theta_R$ which is the sound velocity in the right condensate
[cf. \eqref{vsound}], and the plateau is located between the edges
$z_{0L}\cdot\tau\leq \zeta\leq z_{0R}\cdot\tau$.

\begin{figure}[h]
  \centering
  \includegraphics[width=0.6\linewidth]{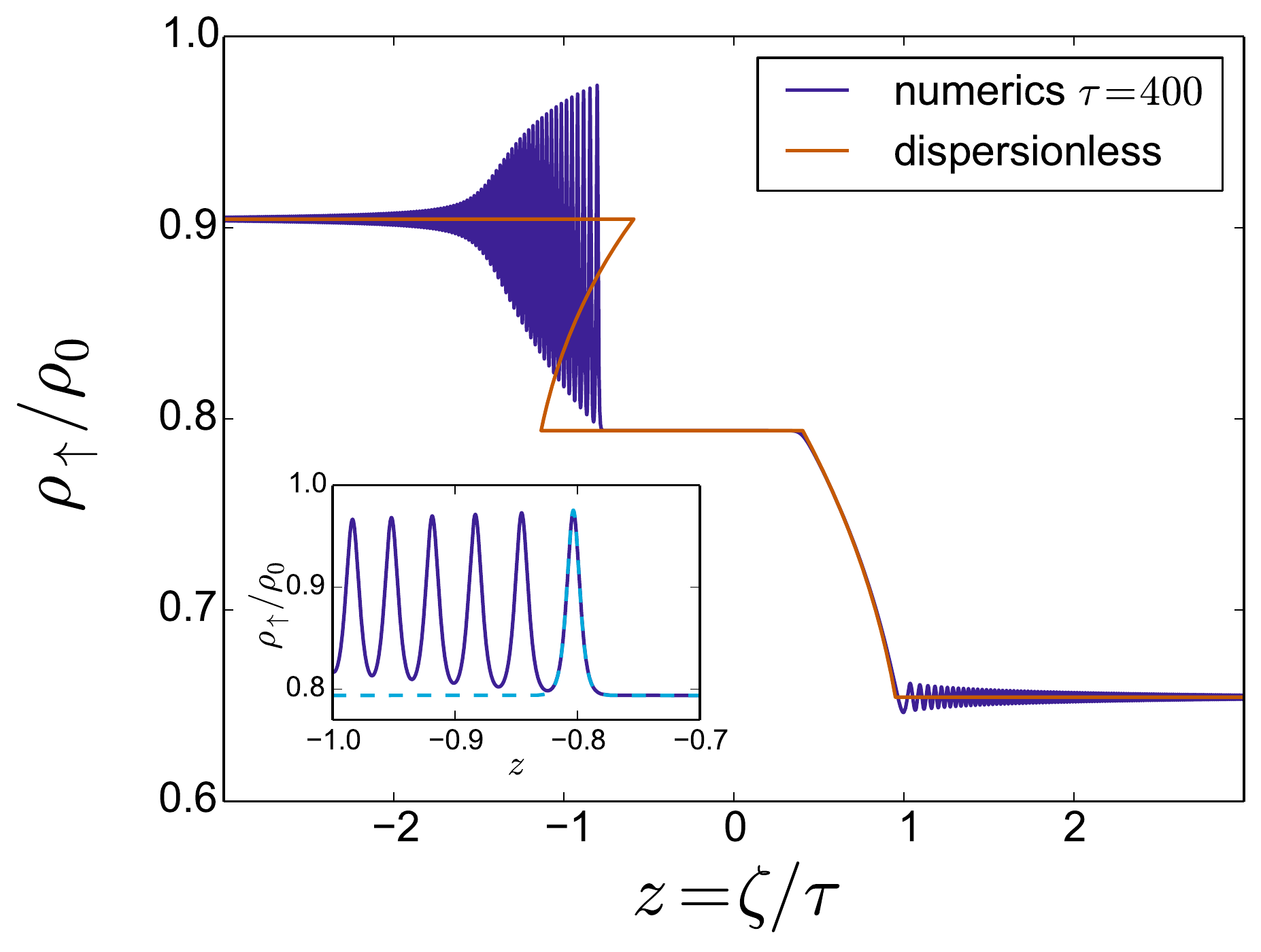}
  \caption{$\rho_\uparrow/\rho_0$ plotted as a function of
    $z=\zeta/\tau$ at $\tau=400$. The initial profile is given by
    \eqref{init-profile} and \eqref{init-profile2}.  The left and
    right asymptotic densities are $\rho_{\uparrow,
      L}/\rho_0=\cos^2(\theta_L/2) = 0.9045$ and $\rho_{\uparrow,
      R}/\rho_0 =\cos^2(\theta_L/2) = 0.6545$.  The dark blue curve
    corresponds to the numerical solution of Eqs.~(\ref{eq8}). The
    orange curve is the result of the dispersionless
    approximation. The inset displays the blow-up of the region of
    the soliton edge of the DSW. The dashed (light blue) line is the
    plot of the first soliton whose characteristics are determined in
    Sec. \ref{sec.WGP}.}\label{fig8}
\end{figure}

Thus, for given values of the densities at both sides of the initial
discontinuity (i.e. for given values of $\theta_L$ and $\theta_R$) one can
calculate the parameters $\theta_0$, $\sigma_0$ defining the plateau
distribution from \eqref{eq57} and determining the ``left'' and ``right''
simple wave solutions joining the quiescent condensates with the
plateau. One of these simple waves represents a rarefaction wave
and the other one describes a formal non-physical multi-valued
solution. This means that the hydrodynamic
approximation fails in the region where the flow is multi-valued
and we have there to take into
account the dispersion effects neglected in the long wavelength
hydrodynamic theory. As a result of dispersion effects, the
multi-valued region is replaced by a dispersive shock wave
which is an oscillatory nonlinear wave structure. Such a situation
is illustrated in Fig.~\ref{fig8}.
There the orange line describes the hydrodynamic approximation
\eqref{eq58}, for which the simple wave at the left of
the plateau is multi-valued. The blue line corresponds to the
numerical solution of the polarization dynamics equations (\ref{eq8})
for an initial profile given by
\begin{equation}\label{init-profile}
 v(\zeta,\tau=0) = 0 \; , \quad\mbox{and}\quad
\theta(\zeta,\tau=0)=
 \tfrac{\displaystyle \theta_R+\theta_L}{\displaystyle 2}+
    \tfrac{\displaystyle\theta_R-\theta_L}{\displaystyle 2}
\tanh\left(\frac{\zeta}{\zeta_0}\right)\; ,
\end{equation}
with
\begin{equation}\label{init-profile2}
\theta_L=0.2\, \pi\; , \quad \theta_R=0.4\, \pi\; ,
\quad\mbox{and}\quad \zeta_0=1\; .
\end{equation}
The value of $\zeta_0$ is not negligibly small, and the argument
previously invoked for justifying the self-similar nature of the flow
does not hold for all times. Instead, the structure of the flow -- with
a well defined plateau joined to both asymptotic regions by specific
structures -- does not appear instantaneously, but takes a finite amount
of time to get formed. As a result, the flow can be considered as
self-similar only for times large compared with this set-up time,
which we numerically evaluate to be of order of $\tau_{\rm set
  up}\simeq 8$ in the case of the initial conditions
specified by \eqref{init-profile} and \eqref{init-profile2}.

It is clearly seen from Fig.~\ref{fig8} that both the right
rarefaction wave and the plateau region are very well described by the
hydrodynamic theory, the dispersion effects leading only to small
oscillations in vicinity of the weak discontinuities located at the
interface between these two regions. On the contrary, the region of
large amplitude oscillations on the left side of the wave pattern is
completely beyond reach of the dispersionless approach and in the next
subsection we shall use a theory able to describe such dispersive
shock wave (DSW) structures with account of dispersion effects.

\subsection{Whitham modulation theory and Gurevich-Pitaevskii problem}\label{sec.WGP}

As seen in Fig.~\ref{fig8}, the numerical solution suggests that the
dispersive shock wave can be seen as a nonlinear periodic solution of
the polarization equations -- such as those studied in Section
\ref{secIII} -- which is however modulated, as shown by the fact that
the amplitude of the oscillations is not constant. This modulation is
gentle, in the sense that the parameters (amplitude, velocity, period,
etc.) of the wave change little over one wavelength and one period of
oscillation. This means that we can apply the Whitham averaging method
for the description of this structure. In his original paper
\cite{whitham-65}, Whitham assumed that the evolution of slowly
modulated nonlinear waves can be described by equations obtained by
averaging the densities and fluxes of the conservation laws over the
rapid oscillations of the wave.  He derived these averaged equations
for several nonlinear wave equations, in particular, for the case of
cnoidal wave solutions of the celebrated Korteweg-de
  Vries (KdV) equation, and---what was most remarkable from a
mathematical point of view---he succeeded in transforming these
equations into a diagonal Riemann form analogous to equations
(\ref{eq45}) obtained in the dispersionless approximation of
hydrodynamic flows. As it became clear later, this success was related
to the specific mathematical properties---complete integrability---of
the KdV equation.

For the case we are interested in, a most important application of the
Whitham theory was suggested by Gurevich and Pitaevskii
\cite{gp-73}. In their approach it was assumed that the expanding DSW
which develops after wave breaking can be described by the nonlinear
periodic solution of the wave equation provided the parameters of this
solution change slowly with time and space coordinate. They
illustrated the method by applying it to the evolution of an initial
step-like discontinuity and to the formation of a DSW after the wave
breaking moment for the KdV wave dynamics.

Since the publications of the work of Whitham and Gurevich and
Pitaevskii, the Whitham theory has been considerably developed in
different directions and has found many applications in nonlinear
physics. In particular, it was shown that many problems can be reduced
to the consideration of the evolution of an initial step-like
discontinuity. It was therefore of great importance to discover
\cite{el} that, for this specific step-like problem, the main
characteristics of DSWs can be obtained by a simple method applicable
to both completely integrable and non-integrable nonlinear wave
equations. In our case the polarization wave dynamics is governed by
the 1D version of the dissipationless 
Landau-Lifshitz equation which is completely
integrable (see, e.g., \cite{kamch-92}). However, the Whitham theory
is not developed well enough for this equation and therefore El's
method \cite{el} seems the most appropriate for the description of the
DSW observed in Fig.~\ref{fig8}.

We thus assume that, instead of the multi-valued solutions found in the
dispersionless approximation in the preceding subsection, a DSW is
generated that joins the neighboring quiescent condensate at the left
side of the wave structure with the plateau region. For definiteness,
and in accordance with the example shown in Fig.~\ref{fig8}, we
consider the case where the Riemann invariant $r_1=\sigma-\theta$ is
constant across the multi-valued region. As was assumed by Gurevich
and Meshcherkin \cite{gm} -- and confirmed in many particular cases --
one of the Riemann invariants preserves its value even after
replacement of the multi-valued solution by the oscillatory DSW: in a
sense, an equality of the type $\left.r_1\right|_-=
\left.r_1\right|_+$ replaces in the case of DSWs the well-known
Rankine-Hugoniot relation of the theory of viscous shocks. It is then
natural to assume that this relation is preserved by the Whitham
averaging method, which yields an appropriate interpolation between
the two edges of the DSW.

As we know, at the small-amplitude edge the DSW can be approximated by
a modulated linear wave (\ref{eq32}), however now propagating along a
non-uniform background corresponding to the simple wave solution with
$r_1=\sigma-\theta=\pi/2-\theta_L=\mathrm{const}$, where we have used
the values of the parameters at the left edge that matches with the
left boundary conditions. With help of this relation we can write
$\sigma=\pi/2-(\theta_L-\theta)$ in the dispersion relation
(\ref{eq15b}), leading to
\begin{equation}\label{eq63}
\Omega(k,\theta)=-\left[2\sin(\theta-\theta_L)\cdot\cos\theta
+ \sqrt{\cos^2(\theta-\theta_L)\sin^2\theta+k^2}\right]k\; .
\end{equation}
In \eqref{eq63} we have chosen the minus sign in front of the square
root of \eqref{eq15b} because we consider wave propagating to the left
with respect to the background condensate.  Equation \eqref{eq63} is the
dispersion of linear waves propagating along a non-uniform
$\theta$-distribution.  During the smooth evolution of the oscillatory
structure the local ``number of waves'' is preserved \cite{Whithbook}
which is expressed by the equation
\begin{equation}\label{nwc}
k_\tau+\Omega_\zeta=0\; .
\end{equation}
Following El \cite{el}, we make a simple-wave type of assumption: in
the DSW the wave number $k$ is a function of $\theta$ only,
$k=k(\theta)$. Then, with account of (\ref{eq63}), the law
\eqref{nwc} of conservation of number of waves can be written under
the form
\begin{equation}\label{eq64}
\frac{d k}{d\theta}\cdot\theta_\tau +
\left(\frac{\prt\Omega}{\prt k}\cdot \frac{dk}{d\theta}
+\frac{\prt\Omega}{\prt \theta}\right)\theta_\zeta=0.
\end{equation}
On the other hand, substitution of $\sigma=\pi/2+\theta-\theta_L$ into
the first of equations (\ref{eq14}) yields
\begin{equation}\label{eq65}
\theta_\tau+
\mathcal{V}\cdot\theta_\zeta=0,
\quad\text{where}\quad \mathcal{V}=
-[2\sin(\theta-\theta_L)\cos\theta+\cos(\theta-\theta_L)\sin\theta].
\end{equation}
Imposing consistency of \eqref{eq64} and \eqref{eq65} considered as
equations for $\theta$, we get
\begin{equation}\label{eq66}
\frac{d k}{d \theta}=
\frac{{\prt\Omega}/{\prt \theta}}{\mathcal{V}-{\prt\Omega}/{\prt k}}.
\end{equation}
This is El's equation that can be extrapolated into the large
amplitude nonlinear region by imposing the condition that the
wavelength tends to infinity at the soliton edge, that is
\begin{equation}\label{eq67}
k=0 \quad \text{at}\quad \theta=\theta_0=(\theta_L+\theta_R)/2.
\end{equation}
Introducing the function
\begin{equation}\label{eq68}
\alpha(\theta)=\sqrt{1+\frac{k^2}{\cos^2(\theta-\theta_L)\sin^2\theta}},
\end{equation}
makes it possible to cast equation (\ref{eq66}) into the form
\begin{equation}\label{eq69}
\frac{d\alpha}{\alpha+1}=
\left(\frac{\sin(\theta-\theta_L)}{\cos(\theta-\theta_L)}-
\frac{\cos\theta}{\sin\theta}\right)d\theta\; ,
\end{equation}
whose solution---with account of the boundary condition (\ref{eq67})---reads
\begin{equation}\label{eq70}
\alpha(\theta)
=\frac{\sin\theta_L+\sin\theta_R}{\cos(\theta-\theta_L)\sin\theta}-1.
\end{equation}
This yields
\begin{equation}\label{eq71}
k(\theta_L)=\sqrt{\sin^2\theta_R-\sin^2\theta_L}.
\end{equation}
Consequently, the left edge of the DSW propagates with
the group velocity evaluated at $k(\theta_L)$:
\begin{equation}\label{eq72}
v_{\mathrm gr}=\left.\frac{\prt \Omega}{\prt k}\right|_{k(\theta_L)}=
-\frac{2\sin^2\theta_R-\sin^2\theta_L}{\sin\theta_R}.
\end{equation}

At the soliton edge of the DSW, we use the ``soliton dispersion law'' \cite{el}
\begin{equation}\label{eq73}
\widetilde{\Omega}(\kappa,\theta)
=-\left[2\sin(\theta-\theta_L)\cdot\cos\theta
+\sqrt{\cos^2(\theta-\theta_L)\sin^2\theta-\kappa^2}\right]\kappa
\end{equation}
relating the velocity $V=\widetilde{\Omega}/\kappa$ of the soliton
with the inverse width $\kappa$ that describes the exponential profile
$w\cong w_3+\frac12(w_4-w_3)\exp\{-\kappa|\zeta+V\tau|\}$ of the soliton
far away from its center (in the regime $|\zeta|\to\infty$). The
relation \eqref{eq73} follows from the remark that the soliton's tail
propagates with the same velocity as the soliton itself and therefore
the soliton's velocity can be found from the asymptotic behavior of
its profile, see, e.g., \cite{ak,dkn}. Again following El, we
assume that along the shock $\kappa=\kappa(\theta)$. Then the
following equation can be derived (see \cite{el}) for this function:
\begin{equation}\label{eq74}
\frac{d \kappa}{d\theta}=\frac{{\prt\widetilde{\Omega}}/{\prt \theta}}
{\mathcal{V}-{\prt\widetilde{\Omega}}/{\prt \kappa}}.
\end{equation}
If we extrapolate the solution of \eqref{eq74} to the small amplitude
region where $\kappa$ tends to zero, we obtain the boundary condition
\begin{equation}\label{eq75}
\kappa(\theta_L)=0.
\end{equation}
Similarly to what has been done for the leading edge of the DSW
[Eq. \eqref{eq66}], it is convenient  for solving Eq.~\eqref{eq74} to
introduce the auxiliary function
\begin{equation}\label{eq76}
\tilde{\alpha}(\theta)=
\sqrt{1-\frac{\kappa^2}{\cos^2(\theta-\theta_L)\sin^2\theta}}.
\end{equation}
Inserting \eqref{eq76} into \eqref{eq74} and taking into account the
boundary condition \eqref{eq75} one obtains
\begin{equation}
\tilde{\alpha}(\theta)=
\frac{2\sin\theta_L}{\cos(\theta-\theta_L)\sin\theta}-1.
\end{equation}
Then, at the soliton edge, $\tilde{\alpha}$ is equal to
\begin{equation}\nonumber
\tilde{\alpha}(\theta_0)=\frac{4\sin\theta_L}{\sin\theta_L+\sin\theta_R}-1,
\end{equation}
and, consequently, this edge propagates with velocity
\begin{equation}\label{sol-vel}
  V_s =\frac{\widetilde{\Omega}(\kappa(\theta_0),\theta_0)}{\kappa(\theta_0)}
=- \frac{1}{2}
(\sin \theta_L + \sin \theta_R) \;.
\end{equation}

\begin{figure}[h]
  \centering
  \includegraphics[width=0.6\linewidth]{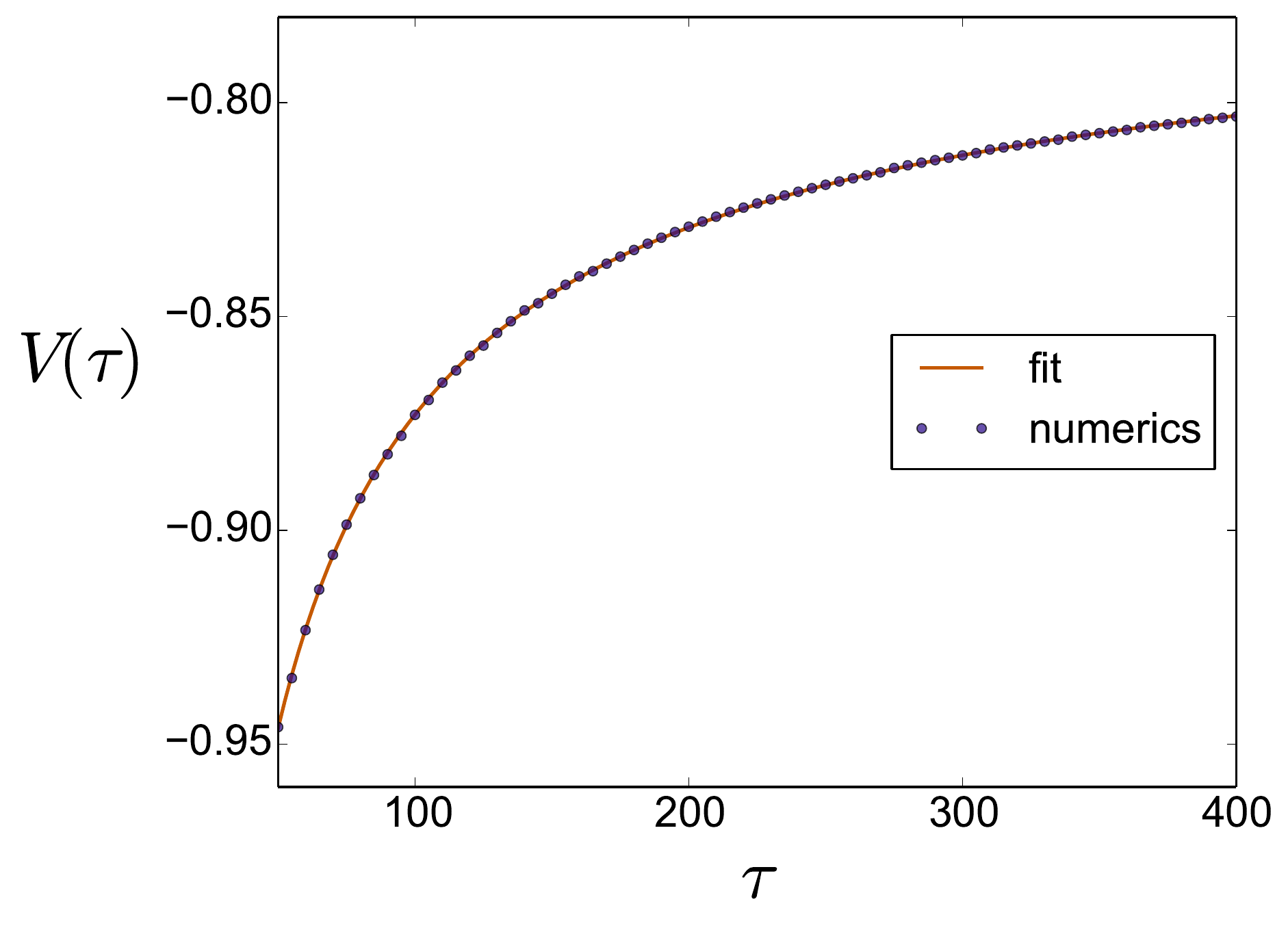}
  \caption{Dots: numerically determined velocity $V(\tau)$ of the
    trailing edge of the numerical solution. The initial conditions
    are specified in Eq. \eqref{init-profile} and
    \eqref{init-profile2}. Continuous line: fit of the numerical datas
    by the formula: $V(\tau)=V_s^{\rm fit}+b \, \tau^{-a}$. One
    obtains $V_s^{\rm fit}=-0.764$, in close agreement with the
    theoretical prediction from Eq. \eqref{sol-vel}: $V_s^{\rm
      theo}=-0.769$.}\label{fig9}
\end{figure}

The comparison of the analytic predictions \eqref{eq72} and
\eqref{sol-vel} for the velocities of the edges of the dispersive
shock wave with our numerical simulations is easily done for the well
defined soliton edge, because, indeed, a leading soliton is easily
identified at this edge of the numerically determined DSW.  The
velocity of this soliton tends for large time to the theoretical
value, as illustrated in Fig.~\ref{fig9}.
In this figure, the numerical result for the velocity $V(\tau)$ of the
soliton at the interface between the DSW and the plateau
region is fitted with the empirical formula $V(\tau)=V_s^{\rm fit} +b
\, \tau^{-a}$, where $V_s^{\rm fit}$, $a$ and $b$ are fitting
parameters. At $\tau=400$, $V$ is still off by about 5\% from its
asymptotic value, but the trend is in excellent agreement with the
prediction \eqref{sol-vel} since one obtains $V_s^{\rm fit} =-0.764$
whereas from \eqref{sol-vel} one expects $V_s^{\rm theo} =
-0.769$. The fitting procedure yields for the other parameters the
values $a=0.74$ and $b=-3.34$. Knowing the velocity of the trailing
edge soliton and the velocity and density of the background plateau
over which it propagates, one can determine from \eqref{eq.w41} all
the parameters $w_1$, $w_2=w_3$ and $w_4$ characterizing the
soliton. Again, the corresponding theoretical profile \eqref{eq31} is
in excellent agreement with the numerics, as shown in the inset of
Fig.~\ref{fig8}. Note that whereas the shape and velocity of the
soliton match the numerics, its position is not exactly the one
expected for a purely self-similar flow (in which case it would be
$z=V_s=-0.769$): this is to be related to the finite set-up time for
creation the flow structure, cf. the discussion presented at the end
of section \ref{hyd-appro} [after Eq.~\eqref{init-profile2}].

As one can see in Fig.~\ref{fig8}, it is difficult from the numerical
solution to unambiguously locate the dispersive edge of the
shock. Hence, at variance with the situation for the soliton edge, the
velocity of the dispersive edge cannot be precisely extracted from the
numerical simulation.  However, one can reasonably argue that the
value $v_{\rm gr}= -1.54 $ obtained from the theoretical formula
(\ref{eq72}) for the initial datas \eqref{init-profile2} matches quite
well with the numerical results (cf. Fig.~\ref{fig8}).

\section{Discussion}\label{sec.discussion}

In this section we discuss the accuracy of the polarization
description of the dynamics of a two-component BEC
[Eqs.~\eqref{eq8}] and also the relevance of our approach to
experimental studies.

\begin{figure}[h]
  \centering
  \includegraphics[width=0.7\linewidth]{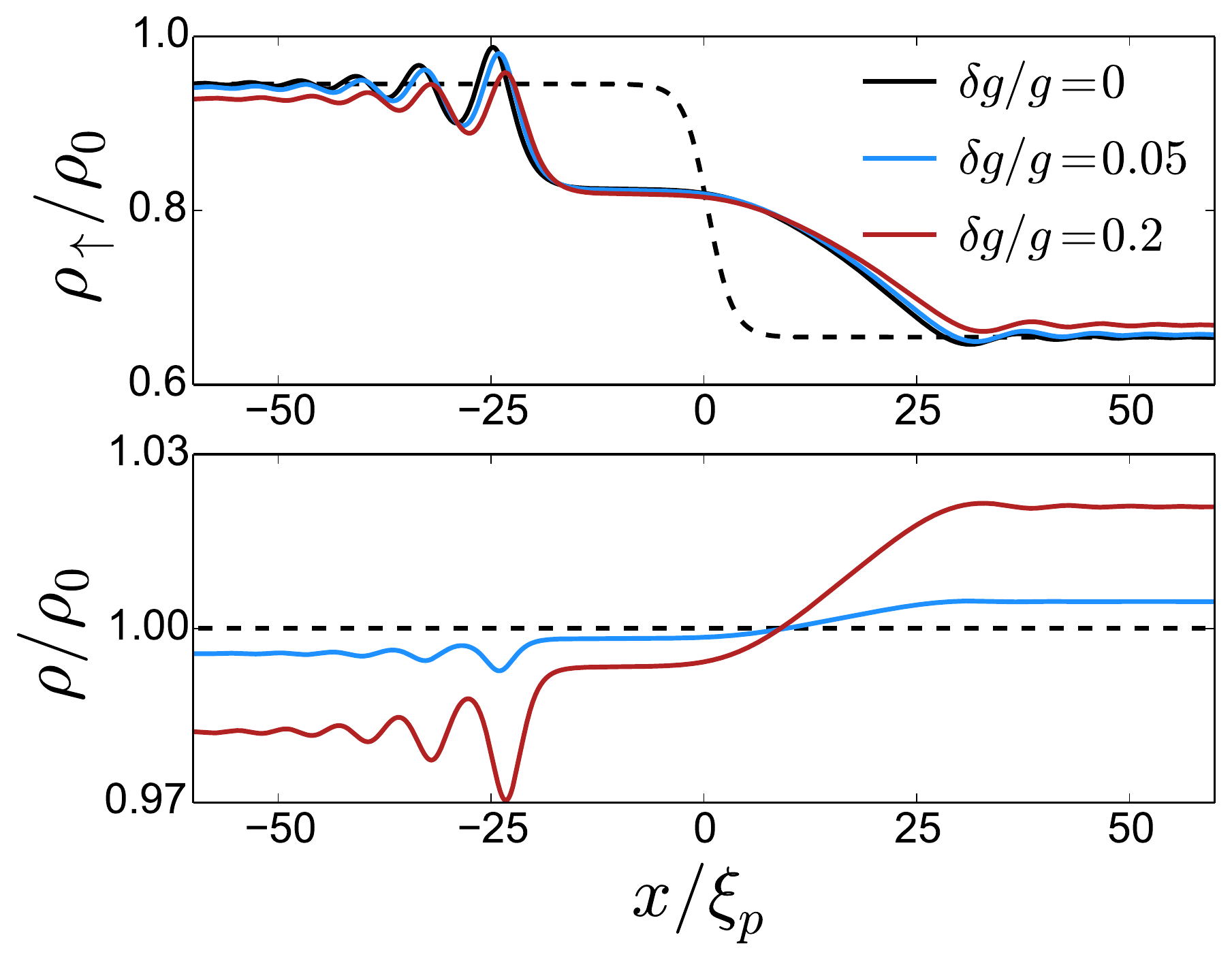}
  \caption{Upper plot: The black solid line represents $\rho_\uparrow$
    as a function of position as obtained from solving the system
    \eqref{eq8} for the initial condition \eqref{init-profile} with
    $\theta_L=0.15\pi$, $\theta_L=0.4\pi$ and $\zeta_0=3$ (dashed
    line). The numerical solution of the Gross-Pitaevskii system for
    the same initial condition and different values of $\delta g/g$ is
    represented by colored lines.  Lower plot: same as above for the
    total density $\rho$.}\label{fig.DSW_test}
\end{figure}

A first question can be asked: in which extend does the assumption of
decoupled dynamics apply? In other words, how small should $\delta
g/g$ be in order for the approach followed in the present work to
apply? A simple way for answering this question is to compare the
results obtained from \eqref{eq8} with the ones obtained from the
numerical solution of the full Gross-Pitaevskii system
\eqref{GP}. This is done in Fig.~\ref{fig.DSW_test} which displays the
evolution of an initial profile of type \eqref{init-profile}. As one
can see from this plot, the agreement is reasonable already for
$\delta g/g=0.2$ and becomes quite good for $\delta g/g=0.05$.  The
lower part of the Figure shows that the assumption of constant total
density is verified with an accuracy of order of $0.5\%$ for $\delta
g/g=0.05$.  We note that the largest departure of the total density
from a constant occurs when $\rho_\uparrow/\rho_0$ is close to unity,
i.e., when $\theta$ is close to 0, as anticipated in Eq.~\eqref{eq6b}.
Note also that the spatial and time scales ($\xi_p$ and $T_p$) are
quite relevant: the Gross-Pitaevskii system is solved for quite
different values of these characteristic scales (the value of $\xi_p$
is multiplied by a factor 2 and the one of $T_p$ by a factor 4 when
one goes from $\delta g/g=0.2$ to $\delta g/g=0.05$), but after the
same time expressed in units of $T_p$ ($24 \, T_p$ in the case of
Fig.~\ref{fig.DSW_test}), the spatial structures almost overlap if the
appropriate units are used.

Another question naturally arises: since Bose-Einstein condensation
of ultra-cold atomic vapors is always realized in trapped systems, it
is important to evaluate the experimental relevance of the
infinitely extended configuration studied in the present work.  One
can first state that the theory has a physical meaning as long as its
characteristic length $\xi_p$ \eqref{eq3}
is much less that the size $X$ of
spatial overlap of the two components which can be estimated in the
framework of the Thomas-Fermi approximation presented in Appendix
\ref{app.TF} :
\begin{equation}\label{cond2}
\xi_p\ll X = \frac{\sqrt{g \rho_0}}{\omega_\parallel \sqrt{m}} \; ,
\end{equation}
where $\omega_\parallel$ is the longitudinal trapping angular
frequency and $\rho_0\simeq N/X$, $N$ being the total
  number of atoms. The condition (\ref{cond2}) combined with
\eqref{cond} reads
\begin{equation}\label{cond3}
\frac{m \omega_\parallel^2 \xi^2}{\rho_0 g} \ll \frac{\delta g}{g} \ll 1\; ,
\end{equation}
where $\xi=\hbar/\sqrt{2 m \rho_0 g}$ is the healing length
($\xi_p=\xi\sqrt{g/\delta g}\,$).  The first inequality of \eqref{cond3}
can be also rewritten as
\begin{equation}\label{cond4}
\omega_\parallel\, \xi\ll c_p \; \quad\mbox{or}\quad
\frac{\xi}{c_p}\ll \frac{1}{\omega_\parallel} \; ,
\end{equation}
that is the polarization sound velocity must be much greater than the
healing length divided by the period of oscillations of atoms in the
trap, or, in other words, the polarization wave passes the healing
length in a time much less that the period of oscillations in the
trap.

It is also worthwhile to address another point: it is known
\cite{Tri00,Toj10,Pol15} that, in the presence of a trapping
potential, the condition of uniform miscibility (which, in our
notations, reads $\delta g>0$) is not sufficient to ensure a good
spatial overlap of the two components. This point is discussed in
Appendix \ref{app.TF} where it is shown that, close to the
mixing-demixing transition, the trapping potential induces a kind of
phase separation if the lower of the intra-species nonlinear constants
(say $g_{\downarrow \downarrow}$) is smaller than the inter-species
constant $g_{\uparrow \downarrow}$, although the criterion of uniform
miscibility $g_{\uparrow \downarrow}<\sqrt{g_{\uparrow
    \uparrow}g_{\downarrow \downarrow}}$ is (weakly) fulfilled.

This phenomenon could explain why, in Ref.~\cite{Dan16}, a kind of
phase separation is observed in the mixture of the two hyperfine
states $|\!\downarrow\rangle = |F=1,m_F=-1\rangle$ and
$|\!\uparrow\rangle=|F=1,m_F=0\rangle$ of $^{87}$Rb in spite of
fulfilment of the uniform mixing condition. For this system
$(a_{\uparrow\uparrow}, a_{\downarrow\downarrow},
a_{\uparrow\downarrow}) = (100.86\,a_0, 100.4\,a_0 ,100.41\,a_0)$,
where $a_0$ is the Bohr radius. Thus $a_{\downarrow\downarrow} <
a_{\uparrow\downarrow}$ which implies mixing of the components in
a uniform case; but non-uniformity caused by the trap potential
induces phase separation. Instead, for the mixture of the two hyperfine
states $|\!\uparrow\rangle = |F=1,m_F=-1\rangle$ and
$|\!\downarrow\rangle = |F=2,m_F=-2\rangle$ of $^{87}$Rb one has
$(a_{\uparrow\uparrow}, a_{\downarrow\downarrow},
a_{\uparrow\downarrow}) = (100.4\,a_0, 98.98\,a_0 ,98.98\,a_0)$, that is the
criterion on miscibility is also fulfilled, but here
$a_{\downarrow\downarrow} = a_{\uparrow\downarrow}$ and the authors
observe a large region of overlap of the two components.

Finally, concerning the comparison of our results with the ones presented in
Ref.~\cite{hceh-11}, it is worth noticing that if $\theta_L\to0$,
that is $\rho_{\uparrow L}\to1$, then the left edge group velocity
\eqref{eq72} tends to the value $v_{\mathrm gr}=-2\sin\theta_R$ which
coincides with the limiting value of velocity (\ref{velo-1}) of the
left edge of the rarefaction wave $z_{0R}$ corresponding to
$\theta_0=0$. This means that the DSW pattern is represented by small amplitude
oscillations around the extrapolation of the rarefaction wave to the region
with $\theta_L\to0$, $\rho_{\uparrow L}\to1$. As a result, the pattern looks like
the rarefaction wave connecting two regions of quiescent condensates with
different values of $\theta$: $\theta_L=0$ and $\theta_R\neq0$.
This apparently agrees with the numerical simulations of the
so-called subcritical regime discussed in \cite{hceh-11} where only
the rarefaction wave was observed for small enough values of the
relative velocity and $\rho_{\uparrow L}=1$.

\section{Conclusion}\label{sec.conclusion}

In vicinity of the mixing/demixing transition, in the limit
\eqref{cond} first identified in Ref.~\cite{qps-16}, the polarization
dynamics decouples from density waves and is described by the
universal equations \eqref{eq8}. In this paper we have identified new
specific polarization structures associated with these equations in
the case of a one dimensional system: algebraic solitons, simple
waves, dispersive shock waves, etc.  But more remains to be done. For
instance, the non-monotonous behavior of the Riemann velocities
(cf. section \ref{sec.simple}) is typically associated to a rich
variety of different types of shocks \cite{kamch-12} which remain to
be investigated in the case at hand; in particular for
  situations with large jumps of the parameter $\theta$, when
DSWs consisting of combined cnoidal and trigonometric
parts are expected. The precise behavior of algebraic solitons in
several instances, and a reliable procedure for their physical
implementation would also be of great interest.  The configuration
described by the initial distributions \eqref{eq55} and \eqref{eq56}
is too schematic for being able to describe the experiments presented
in \cite{hceh-11} where regions with different density ratios are
colliding with finite initial relative velocities. One should thus
consider the case where $\sigma_L$ and $\sigma_R$ are not both equal
to $\pi/2$, and where the plateau formed after the collision is
modulationnally unstable. Finally, the approach developed in this
paper can be generalized to include Rabi coupling between the
components (see, e.g., \cite{Qu16}) and also to two- or
three-dimensional situations \cite{Iac16}. In
particular, formation of oblique polarization solitons by the flow of
the binary condensate past a polarized obstacle (see, e.g.,
\cite{kk-2013}) can be considered in the framework of the present
method.  Works in these directions are in progress.

\section*{Acknowledgements}
  We thank S. Stringari for fruitful discussions. AMK thanks
  Laboratoire de Physique Th\'eorique et Mod\`eles Statistiques
  (Universit\'e Paris-Sud, Orsay) where this work was started, for
  kind hospitality.

\paragraph{Funding information}
This work was supported by the French ANR under
  grant n$^\circ$ ANR-15-CE30-0017 (Haralab project).

\begin{appendix}

\section{Computation of the energy of a soliton}\label{app.energy}

We briefly present here the computation leading to the result
\eqref{eq.ener.sol} for the energy of the soliton. From
\eqref{pt.traveling} and \eqref{tildephi} one gets $\phi_\zeta=V
(B-w)/(1-w^2)$ with $B=(1-w_2^2)v_0/V+w_2$ and from \eqref{eq16}
$\theta_\zeta^2=-Q(w)/(1-w^2)$. This yields for the
energy \eqref{grand.pot.sol}
\begin{equation}\label{eq.a1}
{\cal E}=\int_{\mathbb{R}}\frac{d\zeta}{2}
\left\{\frac{-Q(w)}{1-w^2} +(1-w^2)
\left[V^2\frac{(B-w)^2}{(1-w^2)^2}-1\right]
+(1-v_0^2)(1-w_2^2)\right\}\; .
\end{equation}
The integrand being symmetric ---since $w(\zeta)$ is--- one can thus
restrict the range of integration to the domain $(-\infty,0]$ over
which one can write $d\zeta=+ dw/\sqrt{-Q(w)}$. Using the fact that
one can express $B$, $v_0$ and $V$ as functions of $w_1$, $w_2$ and
$w_4$ [cf. Eq. \eqref{eq.w41}], it is then possible to re-write
\eqref{eq.a1} under the form
\begin{equation}\label{eq.a2}
{\cal E}=\int_{w_2}^{w_4} dw \,  \frac{2w-w_2-w_4} {\sqrt{(w_4-w)
(w-w_1)}} \;,
\end{equation}
which yields the result \eqref{eq.ener.sol}.

\section{Effective demixing in a 1D trap}\label{app.TF}

In this appendix we present 1D computations in the framework of the
Thomas-Fermi description of the system \eqref{GP} in the presence of a
trapping potential \cite{Ho96}. It is known \cite{Pu98} that the
Thomas-Fermi approximation cannot quantitatively describe all the
possible configurations encountered the mixture of two BECs, but it
will permit to identify specific situations which will then have to be
confirmed by a full numerical solution.

We consider here $N_\uparrow$ and $N_\downarrow$ atoms of each
component placed in a harmonic potential of longitudinal angular
frequency $\omega_\parallel$ much smaller than the radial trapping
angular frequency $\omega_\perp$. In the so called ``1D mean field
regime'' \cite{Men02}, the system can be described by the effective 1D
Gross-Pitaevskii equation \eqref{GP} with $g_{\uparrow \uparrow}=2
\hbar\omega_\perp a_{\uparrow \uparrow}$ \cite{Olsh98} where
$a_{\uparrow \uparrow}$ is the 3D intra-species $s$-wave scattering
length of the ``up'' component (an similar expressions for
$g_{\downarrow \downarrow}$ and $g_{\uparrow \downarrow}$). In the
situation we are interested in where $N_\uparrow\sim N_\downarrow$ and
$a_{\uparrow \uparrow}\sim a_{\uparrow \downarrow} \sim a_{\downarrow
  \downarrow}$, the 1D mean field regime holds when
$N_\uparrow(\omega_\parallel/\omega_{\perp})(a_{\uparrow\uparrow}/a_\perp)\ll
1$, where $a_\perp=\sqrt{\hbar/m\omega_\perp}$ is the radial harmonic
oscillator length.

\begin{figure}
\begin{center}
\includegraphics[width=0.495\linewidth]{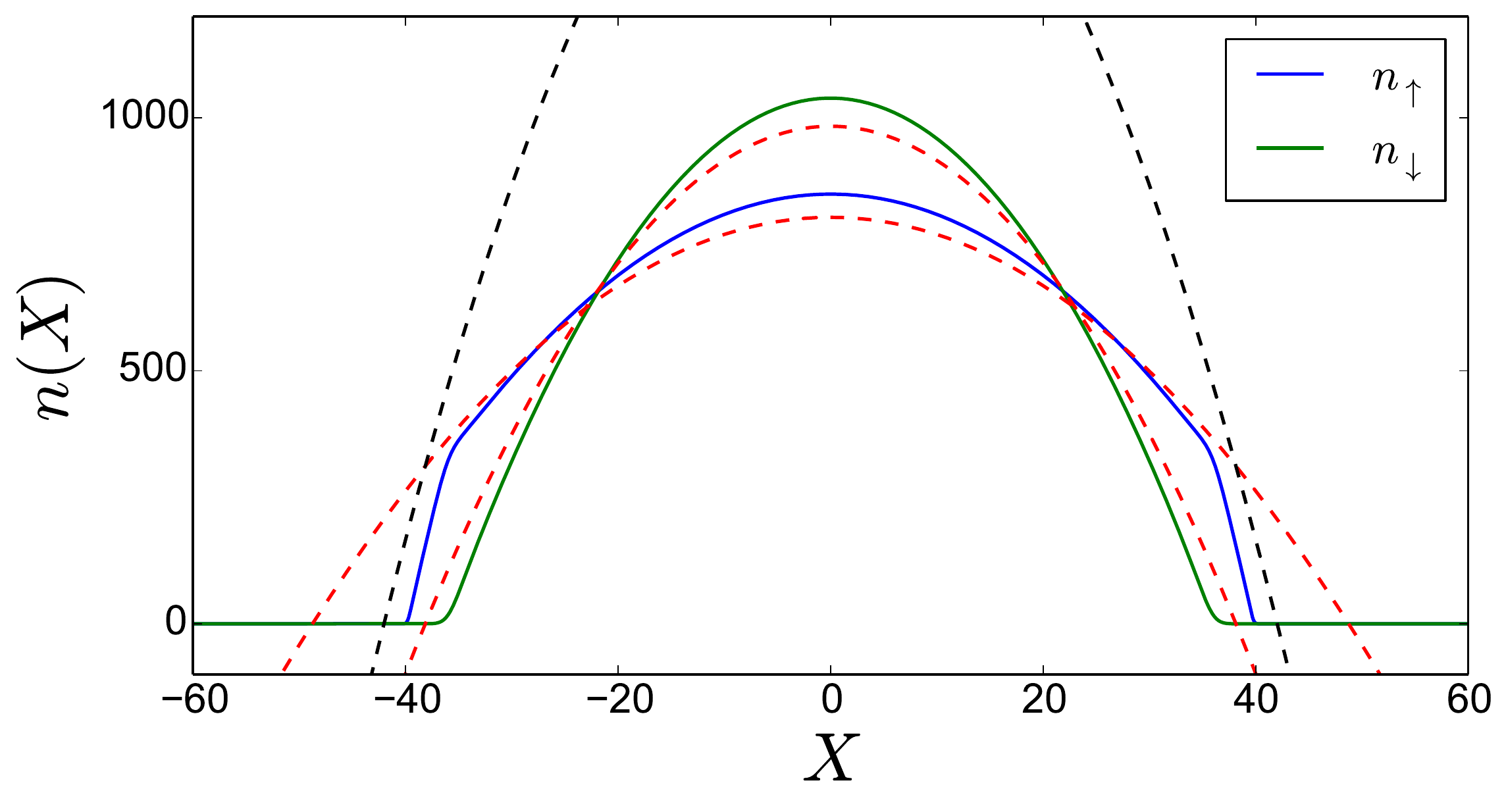}
\includegraphics[width=0.495\linewidth]{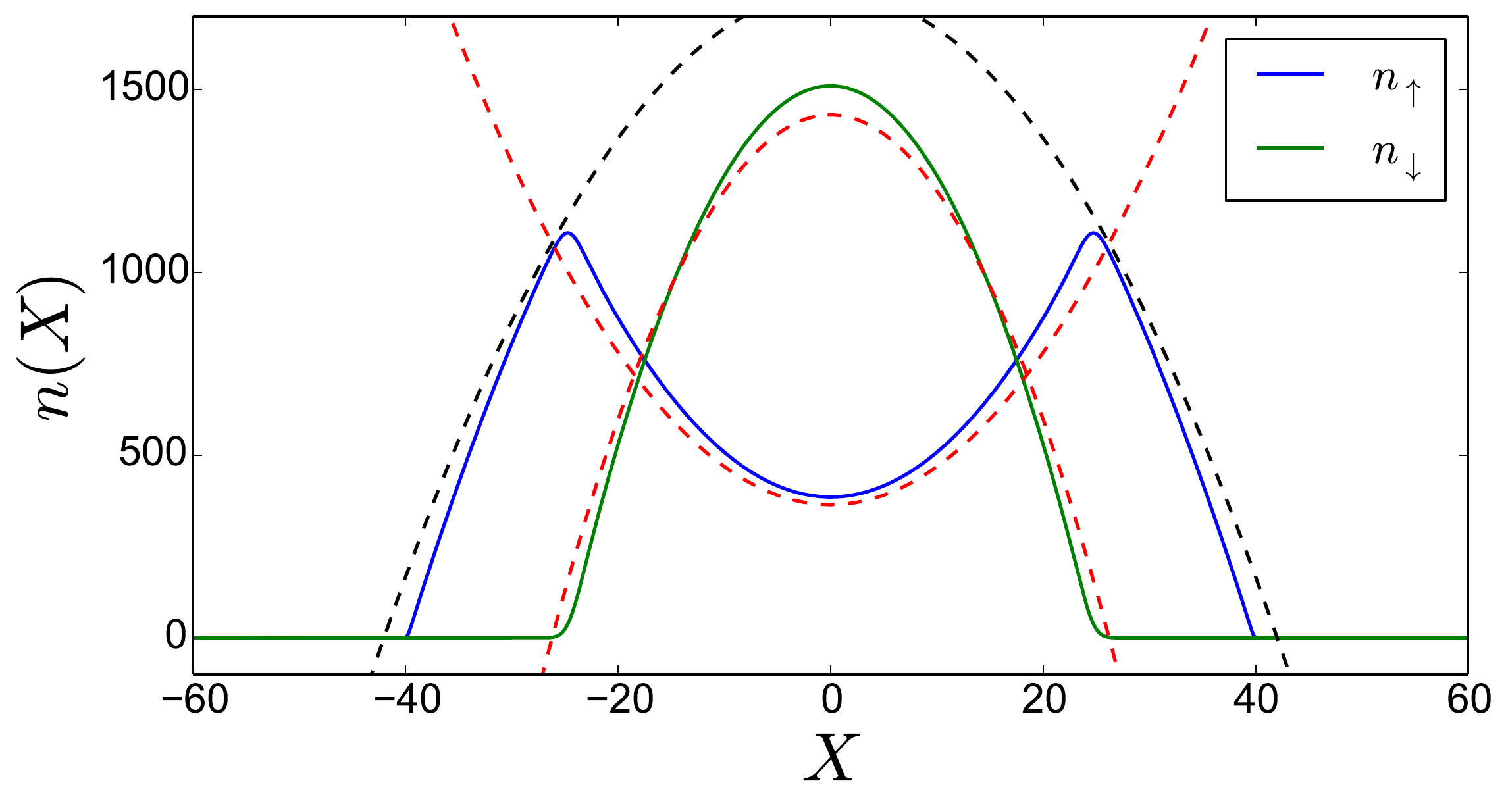}
\end{center}
\caption{Distribution of atoms in a 1D trapped two-component BEC. The
  trap parameters are $\omega_\parallel=2\pi\times1 \,\rm Hz$,
  $\omega_\perp = 2\pi\times500 \, \rm Hz$ and $N_\uparrow =
  N_\downarrow = 5\times 10^4$. The condensates are formed by
  $^{87}$Rb atoms, this yields $a_\parallel=10.8$ $\mu$m. The red
  dashed lines correspond to $n_\uparrow^a(x)$ and $n_\downarrow(x)$,
  Eqs. \eqref{eq.b6} and \eqref{sol:n2}. The black dashed lines
  display $n_\uparrow^b(x)$ \eqref{eq.b7}. The solid lines correspond
  to the numerical solution of the Gross-Pitaevskii equations. The
  left plot corresponds to the values $(a_{\uparrow\uparrow},
  a_{\downarrow\downarrow}, a_{\uparrow\downarrow}) = (102\,a_0,
  101\,a_0 ,100\,a_0)$ for the scattering lengths. The right plot
  corresponds to $(a_{\uparrow\uparrow}, a_{\downarrow\downarrow},
  a_{\uparrow\downarrow}) = (102\,a_0, 99\,a_0 ,100\,a_0)$. The
  precise values of these scattering lengths have been chosen for
  exemplifying the phenomenon of effective demixing, but they all lie
  within a realistic range for $^{87}$Rb.}\label{fig.cases}
\end{figure}

We chose the parameters so that the mean field condition of
miscibility $\sqrt{a_{\uparrow \uparrow} a_{\downarrow\downarrow}} >
a_{\uparrow \downarrow}>0$ is always fulfilled, and in the following
we denote as $A$ the parameter having the dimension of length defined
by
\begin{equation}\label{eq.b1}
A^2 = a_{\uparrow\uparrow}a_{\downarrow\downarrow} -
a_{\uparrow\downarrow}^2 >0\; .
\end{equation}
We define the non-dimensional position $X=x/a_\parallel$, where
$a_\parallel=\sqrt{\hbar/m\omega_\parallel}$ is the longitudinal
harmonic oscillator length, and the non-dimensional densities
$n_{\uparrow,\downarrow}$ such that $\int n_{\uparrow,\downarrow}(X)
dX=N_{\uparrow,\downarrow}$. We denote as ``down'' the component for
which the intra-species interaction is the lowest, i.e., $a_{\downarrow
  \downarrow}<a_{\uparrow \uparrow}$.  Within the Thomas-Fermi
approach one obtains
\begin{equation}
\label{sol:n1}
n_\uparrow(X) =
\left\{
\begin{array}{ll}\displaystyle
n_\uparrow^a(X) & \mbox{if}\; |X| \leq X_\downarrow, \\
n_\uparrow^b(X) & \mbox{if}\; X_\downarrow \leq |X| \leq
X_\uparrow, \\
0 & \mbox{if}\; X_\uparrow \leq |X|,
\end{array}
\right.
\end{equation}
and
\begin{equation}\label{sol:n2}
n_\downarrow(X) =
\left\{
\begin{array}{ll}
\frac{\omega_\parallel}{\omega_\perp}
\frac{(a_{\uparrow\uparrow}-a_{\uparrow\downarrow})
a_\parallel}{4\,A^2} \left(X_\downarrow^2-X^2 \right)& \mbox{if}\; |X| \leq
X_\downarrow, \\
0 & \mbox{if}\; X_\downarrow \leq |X|,
\end{array}
\right.
\end{equation}
where
\begin{equation}
X_\uparrow^3 = \frac{3\,\omega_\perp}{\omega_\parallel}\,
\frac{a_{\uparrow\uparrow}N_\uparrow+a_{\uparrow\downarrow}N_\downarrow}{a_\parallel}
\;,\quad
X_\downarrow^3 = \frac{3\,\omega_\perp}{\omega_\parallel}\,\frac{A^2
N_\downarrow}{(a_{\uparrow\uparrow}-a_{\uparrow\downarrow})a_\parallel} \;,
\end{equation}
\begin{equation}\label{eq.b6}
n_\uparrow^a(X)= \frac{\omega_\parallel}{\omega_\perp}
\left[\frac{a_\parallel}{4\,a_{\uparrow\uparrow}}X_\uparrow^2 -
\frac{a_{\uparrow\downarrow}a_\parallel}{4\,A^2}\left(1-
\frac{a_{\uparrow\downarrow}} {a_{\uparrow\uparrow}} \right)
X_\downarrow^2  -
\frac{(a_{\downarrow\downarrow}-a_{\uparrow\downarrow})a_\parallel}{4\,A^2}
\, X^2 \right]\; ,
\end{equation}
and
\begin{equation}\label{eq.b7}
n_\uparrow^b(X)= \frac{\omega_\parallel}{\omega_\perp} \,
\frac{a_\parallel}{4\,a_{\uparrow\uparrow}}
\left(X_\uparrow^2-X^2 \right) \; .
\end{equation}
These results are compared in Fig.~\ref{fig.cases} with the numerical
solutions of Eqs.~\eqref{GP} in the presence of a trapping potential
$V(x)=\tfrac12 m \omega^2_\parallel x^2$. The two plots of this figure
are drawn for a configuration verifying the miscibility condition
\eqref{eq.b1} \footnote{For the chosen sets of parameters, one is at
  the limit of the 1D mean field regime : $N_\uparrow
  (\omega_\parallel/\omega_{\perp}) (a_{\uparrow\uparrow}/a_\perp)
  \simeq 1$. The condition of applicability of the Thomas-Fermi
  approximation \cite{Men02} is well fulfilled: $[N_\uparrow
  (a_{\uparrow\uparrow}/a_\perp)
  \sqrt{\omega_\perp/\omega_\parallel}\,]^{1/3} \simeq 23 \gg 1$.}. In
the left plot $a_{\downarrow\downarrow} > a_{\uparrow\downarrow}$
whereas the situation is reversed in the right one (similar plots have
already been obtained in Ref.~\cite{Tri00}).
Although the corresponding change of scattering lengths is minute,
close to the mixing-demixing transition the effect is spectacular: one
reaches a situation of quasi-demixing where the component with the
largest scattering length (the up component)
is expelled from the trap's center. This
situation would be expected in the situation $a_{\downarrow\downarrow}
\ll a_{\downarrow\uparrow} \simeq a_{\uparrow\uparrow}$. The point is here
that the same effect is observed for a system verifying the
miscibility condition \eqref{eq.b1} provided one remains close to
immiscibility and that $a_{\downarrow\downarrow} \lesssim
a_{\downarrow\uparrow}$. The parameter governing the expulsion of the
up component from the center of the trap is the non-dimensional
curvature of its density at $X=0$. From \eqref{eq.b6} this parameter is equal
to
\begin{equation}
  -\frac{\omega_\parallel}{\omega_\perp}\times
  \frac{\left(a_{\downarrow\downarrow} - a_{\uparrow\downarrow}\right)a_\parallel}
{a_{\uparrow\uparrow}a_{\downarrow\downarrow} -
    a_{\uparrow\downarrow}^2} \; .
\end{equation}
In the cases presented in Fig.~\ref{fig.cases} the value of this
parameter changes from $-1.3$  (in the left plot of the figure) to
$+4.2$ (right plot) just by changing $a_{\downarrow\downarrow}$ by 2\%.
\end{appendix}

\end{document}